\documentclass{ws-ijmpb}
\usepackage{subfigure}
\begin{document}
\newcommand \be {\begin{equation}}
\newcommand \ee {\end{equation}}
\newcommand \bea {\begin{eqnarray}}
\newcommand \eea {\end{eqnarray}}
\newcommand \nn {\nonumber}
\newcommand \la {\langle}
\newcommand \ra {\rangle}
\newcommand \ve {\varepsilon}
\newcommand{\dd}[0]{\mathrm{d}}
\newcommand{\Tr}{\mathbf{\mathfrak{Tr}}}
\newcommand{\prob}[0]{\mathrm{Prob}}
\newcommand{\LL}{\textsf{L}}
\newcommand{\Tet}{T^{*}(L)}
\newcommand{\PDF}{\Pi}
\newcommand{\Dint}{{\cal D}}
\newcommand{\ff}[2]{\frac{#1}{#2}}
\newcommand{\D}{\mathcal{D}}
\newcommand{\Tc}{T_\mathrm{c}}
\newcommand{\qv}[0]{{\bm{q}}}
\newcommand{\action}{\mathcal{S}}
\newcommand{\ttha}{\mathcal{T}}

\markboth{Maxime Clusel, Eric Bertin}
{Global fluctuations in physical systems}

%
\catchline{}{}{}{}{}
%

\title{GLOBAL FLUCTUATIONS IN PHYSICAL SYSTEMS: A SUBTLE INTERPLAY BETWEEN
SUM AND EXTREME VALUE STATISTICS}

\author{MAXIME CLUSEL}

\address{ 
Department of Physics and Center for Soft Matter Research,
New York University,
\\ 4 Washington Place, New York, NY 10003, United States of America\\
maxime.clusel@nyu.edu}

\author{ERIC BERTIN}

\address{Universit\'e de Lyon, Laboratoire de Physique,\\
Ecole Normale Sup\'erieure de Lyon, CNRS,
46 all\'ee d'Italie, F-69007 Lyon, France\\
eric.bertin@ens-lyon.fr}

\maketitle

\begin{history}
\end{history}

\begin{abstract}
Fluctuations of global additive quantities, like total energy or magnetization
for instance, can in principle be described by statistics of sums of
(possibly correlated) random variables.
Yet, it turns out that extreme values (the largest value among a set
of random variables) may also play a role in the statistics of
global quantities, in a direct or indirect way.
This review discusses different connections that may appear
between problems of sums and of extreme values of random variables, and
emphasizes physical situations in which such connections are relevant.
Along this line of thought, standard convergence theorems for sums and
extreme values of independent and identically distributed random variables
are recalled, and some rigorous results as well as more heuristic reasonings
are presented for correlated or non-identically distributed random variables.
More specifically, the role of extreme values within sums of broadly
distributed variables is addressed, and a general mapping between
extreme values and sums is presented, allowing us to identify
a class of correlated random variables whose sum 
follows (generalized) extreme value distributions.
Possible applications of this specific class of random variables are
illustrated on the example of two simple physical models.
A few extensions to other related classes of random variables sharing
similar qualitative properties are also briefly discussed, in connection
with the so-called BHP distribution.
\end{abstract}


\section{Introduction}

The present review deals with the issue of
fluctuations of global quantities
and the possible relevance of extreme values in the statistics of these
fluctuations. A physical quantity is called a global quantity,
or global observable, when it is defined as the spatial sum of local
quantities in a large enough system (or subsystem).
For instance, the total energy and the
total magnetization of a macroscopic system are global observables.
Such quantities are very frequent in physics since on the one hand
they are useful characterizations of macroscopic systems, and on the other
hand most measurement apparatus have a resolution that is much larger than
the scale of individual degrees of freedom. Hence from a practical
viewpoint, most measured quantities turn out to be sums of local quantities,
and thus fall into the category of global observables.

When measuring a global observable, one usually records a time signal, and
the first quantity that can be determined from this signal is
the mean value, which is a natural characterization of the system (assuming
the system to be in a stationary state; extension to slow time evolution
may also be considered). Then, to go beyond mean values, it is interesting
to quantify fluctuations around the mean, for instance by determining the
empirical variance of the signal.
A more refined information is given by the full probability
distribution of the fluctuations.
Such a distribution is of particular interest since
it may give information on the physics of the system. The main reason
for this is the existence of the Central Limit Theorem (CLT), which states
that a sum of a large number of independent and identically distributed
(i.i.d.) random variables is distributed according to a Gaussian (or normal)
distribution, provided that the second moment of each individual variable
is finite --a more precise statement of the theorem will be given below.

The CLT, as any other theorem, relies on some assumptions that are of
course not always true in physical systems. 
In particular, the second moment of the individual variables may not
be finite, in which case other different asymptotic distributions
(the so-called L\'evy-stable laws) are reached in the limit of an
infinite number of terms. This may happen in different physical contexts,
like glasses and disordered
systems,\cite{Mon73,Shl74,Mon75,Shl88,BG,Bouchaud92,Bouchaud95}
laser cooling experiments,\cite{BBE94,BBA02}
turbulent flows,\cite{SZK93,SWS93,Min96} or
blinking of nanocrystals,\cite{Brokmann} to quote only a few of them.
Qualitatively, the departure from the Gaussian distribution is
in this case due to the fact that a few terms
become extremely large and dominate the sum. In contrast, when the
second moment is finite, all terms within the sum are of the
same order, and none of them play a dominant role.

Another situation of physical relevance, for which Gaussian distributions
may not be found (even in the limit of an infinite number of terms),
is when the variables that are summed are correlated.
Intuitively, one may expect that the Gaussian distribution still holds
when the correlations are ``weak'', so that ``strong'' enough correlations
may be necessary to prevent the distribution of the sum from converging
to the Gaussian law. On the mathematical side, one must admit that
a rigorous generic approach to this problem is a really difficult task
and, to the best of our knowledge, no general results or criteria
seem to be available.
Hence correlated variables can only be treated case by case,
leading to a (probably still incomplete) gallery of various limit
distributions.

From a physicist viewpoint, a qualitative understanding of systems
involving sums of correlated variables may be gained
in some cases from simple scaling arguments, allowing one to guess
whether the CLT (or an extension to it) is expected to hold or not.
The simplest of these arguments is an estimate of the number
of ``effectively independent'' random variables. When this number
diverges with the total number of terms in the sum, it is reasonable
(at least to the physicist's mind) to consider that the limit distribution
should be Gaussian (if the second moment is finite). 
Still, a significant number of physical systems do not fulfill this
condition, and fluctuations of global observables then display a
non-Gaussian statistics, even in the infinite size limit.
Indeed, a vast number of different non-Gaussian distributions have
been reported in different contexts, like critical
phenomena,\cite{Bruce81,Binder81,Binder92,Zheng01,Zheng03}
width distribution of growing
interfaces,\cite{Plischke94,Racz94,Verma00,Marinari02,Rosso03,Moulinet04}
turbulent flows,\cite{Frisch95,Peinke97,Carreras99,Portelli03,Chevillard05}
driven nonequilibrium systems,\cite{Aumaitre01,Farago02,Droz03,Visco06}
or different types of quantum systems,\cite{Weill05,Pilgram03,Reulet03}
to give only a few examples. It is interesting to note that
among these non-Gaussian distributions, asymmetric
distributions rather similar to those appearing in the context of extreme value
statistics (the statistics of the maximum or minimum value in a set
of random variables) have been repeatedly found.\cite{BHP,Pinton99,Bramwell01,Bramwell01b,Noullez02,Holdsworth02,Tothkatona03,Fedorenko03,Pennetta04,Chamon04,Wijland04,Vanmil05,Brey05,Duri05,Varotsos05}
It is thus natural to wonder whether this is a mere coincidence,
or if there may be some generic reasons for such a similarity.

Altogether it turns out that, in a rather unexpected way, extreme values 
seems to play a role in the statistics of random sums, both for
sums of broadly distributed random variables and for sums of
correlated or non-identically distributed variables.
The analysis of these two problems actually reveals that the role
of extreme values is very different in both cases. Specifically, in the
case of correlated variables, the similarity with extreme value statistics
does not come from a dominant contribution of the largest terms, but rather
from a natural, though not obvious, mapping between extreme values
and sums of random variables.

Having this intricate picture in mind, we believed it would be
relevant to treat essentially on the same footing the statistics
of sums and of extreme values of random variables, before dealing
with the different relationships arising between these fields.
Accordingly, the paper is organized as follows.
Some elementary statistical concepts are briefly recalled
in Sect.~2. Standard mathematical theorems concerning the convergence
to asymptotic distributions of both random sums and extreme values
are presented in Sect.~3, in the case of independent and identically
distributed variables. The case of dependent and of
non-identically distributed variables is addressed in Sect.~4,
paying particular attention to physical arguments and examples.

Let us emphasize that giving an exhaustive review of such a large
topic as sums and extreme values of random variables, is obviously
far beyond the scope of the present review paper. Accordingly, the latter
has the more modest aim of giving the reader an overall
(though incomplete) picture, from a physicist perspective,
of the basic results and issues in these fields, with specific emphasis
on some recent results providing unexpected connections between them.
In this spirit, we tried to make mathematical statements
as precise as possible, while also leaving room for numerous 
physical discussions and examples.

\section{Some probabilistic concepts useful in physics}

\subsection{Random variables: dependence, joint probability and spatial
organization}
\label{rv-ijp}

One of the main goals of statistical physics is to study the macroscopic
properties of models defined by some microscopic interaction rules
between "particles", or more generally, microscopic degrees of freedom.
The paradigm behind this approach is the belief that there exists,
at least for some classes of systems,
general mechanisms for going from the microscopic rules to a
macroscopic behavior, so that the knowledge gained from
the study of specific models goes beyond a collection
of particular results. In this section, we would like to go from the
physicist's viewpoint to the mathematician's one: this will be useful
in the rest of the present article, and might also illustrate the way
mathematics and physics interact in statistical mechanics.

Let us start by one of the archetypal model of statistical physics,
namely the Ising model.
On each site $i=1,\ldots,N$ of a $D$-dimensional square
lattice,\footnote{Throughout the article, we generically denote as $N$
the number of random variables considered.} a spin variable can
take two possible values $s_i=\pm 1$. Each spin $s_i$
interacts with its nearest neighbor and with an external magnetic field
$h$, so that the total energy of a configuration
$s \equiv \{s_i\}_{i=1 \ldots N}$ is given by the celebrated Hamiltonian:
\be
\mathcal{H}[s]=-K\sum_{\langle i,j \rangle} s_i s_j -
h\sum_{i=1}^N s_i, \qquad s \equiv \{s_i\}_{i=1 \ldots N}.
\ee
The summation in the first term is over all pairs of nearest neighbor sites
$i$ and $j$. A global observable a statistical physicist may be
interested in is the magnetization $m[s]$ of a given configuration:
\be
m[s]=\frac{1}{N}\sum_{i=1}^N s_i.
\ee
In the canonical ensemble, the average of this quantity at a given inverse
temperature $\beta=1/k_B T$, is defined by
\be
\la m \ra = \sum_{s} P[s] m[s],\ \quad
P[s]=\frac{1}{\mathcal{Z}} e^{-\beta \mathcal{H}[s]}.
\ee
The normalization constant $\mathcal{Z}$ is the canonical partition
function of the model. The function $P[s]$ is then the probability
of a given configuration $s$.
Most often, it is not an easy task to explicitly compute the mean
magnetization. The problem comes from the nearest neighbor
interaction, which induces statistical correlations between sites $i$
and $j$.
Spins are independent only in the specific case $K=0$, when no interactions
are present. In this case one easily obtains
$\la m \ra = \mathrm{tanh}(\beta h)$.

The above physical example could be rephrased in a more mathematical
language.
In probabilistic terms, a spin configuration $s$ forms a set of random
variables $(s_1,\ldots,s_N)$, and the probability $P[s]$ of
a configuration is denoted as the joint probability
$J_N(s_1, \ldots, s_N)$ of the random set.
The random set is fully characterized by its joint probability,
which is the probability
to obtain the particular set $(s_1,\ldots,s_N)$ from a
realization of the random variables. From this joint probability one
can define the marginal probability density $p_i(s_i)$
associated to the random variable $s_i$, by summing $J_N(s_1, \ldots, s_N)$
over all the remaining variables:
\be
p_i(s_i) = \sum_{\{s_j=\pm 1, j\ne i\}}
J_N(s_1,...,s_{i-1},s_i,s_{i+1},...,s_N).
\ee
The $N$ random variables $s_i$
are independent if and only if the joint probability can be factorized
as the product of the marginal distributions $p_i(s_i)$, 
\be \label{defind}
J_N(s_1,...,s_N)=\prod_{i =1}^N p_i(s_i).
\ee
In the Ising model, and more generally in canonical equilibrium systems,
there is an obvious link between the joint probability and the Hamiltonian:
\be
J_N(s_1,...,s_N)=\frac{1}{\mathcal{Z}}
\exp \left[ -\beta \mathcal{H}(s_1,...,s_N)\right],
\ee
and the non-interacting case ($K=0$) corresponds to
independent variables in the probabilistic view:
\bea
J_N(s_1,...,s_N)=\frac{1}{\mathcal{Z}} \prod_{i=1}^N
e^{-\beta h \sigma_i}.
\eea

This elementary example shows how problems of statistical physics can be
recast into a probabilistic language. Let us point out that
in physics, in contrast to what happens in mathematics, a meaning is given
to the random quantities, and to their labels.\cite{Feynman}
It introduces some key
concepts from the physicist's viewpoint such as a distance (spatial or
temporal) and a space dimensionality.  This is the reason why the description
of the dependence of random variables in terms of the joint probability
is somehow too rich for the physicist, who often prefers a simpler
characterization through the two-point correlation function:
\bea \label{corrfunc}
C_{ij}=\la s_{i}s_{j} \ra-\la s_{i}\ra \la s_{j} \ra,
\eea
which emphasizes the spatial structure of the model.
Under the assumption of homogeneity and isotropy,
an even more 'compact' information can be obtained from
a correlation length or time, defined
as the characteristic scale of the correlation function (\ref{corrfunc}),
\be
C_{ij} = C_0 f\left( \frac{r_{ij}}{\xi}\right),
\ee
where $f$ is a dimensionless function and $r_{ij}$ is the euclidean
distance between sites $i$ and $j$.
Quite often, a lot of information on a physical system can be gained
from the behavior of the correlation length with temperature, magnetic
field or other control parameters.

\subsection{Concept of asymptotic distributions}
\label{sect-asympt-dist}

As already pointed out, the specificity of physics (and of other sciences
aiming at describing real systems in a mathematical language)
is to give an interpretation, in connection with the real world, to
the mathematical objects involved in the proposed description.
From the physicist's point of view, probability theory provides various
theorems that are kinds of ``reasoning shortcuts'', ready to be used in a
given physical context. Among the most useful theorems are the
convergence theorems, which will be presented in Sect.~\ref{sect-convergence}.
In the present section, we recall elementary definitions and probabilistic
tools. Our aim is not to present in a rigourous way probability
concepts but to introduce practical tools that are needed in the
following sections.

We already considered discrete random variables, like spin variables $s$,
in the previous section. Discrete random variables are most often
characterized by the probability $P[s]$ of each configuration $s$.
In contrast, when considering continuous random variables, several
probabilitic tools may be used depending on the context.
From the physicist's point of view, a continous random
variable\footnote{In the rest of the paper, we shall most of the time use
the same notation for a random variable $X$ and its value $x$.} $X$
is characterized by a non-negative function $p(x)$, the probability
density function (PDF), or probability distribution, of $X$.
This PDF is such that the cumulative
probability distribution $F(x)$ can be expressed as
\be
\prob(X\le x) \equiv F(x)=\int_{-\infty}^x \dd x'\; p(x').
\ee
The random variable $X$ can equivalently be described by its characteristic
function $\chi(\omega)$, defined as the Fourier transform of $p(x)$:
$$
\chi (\omega)=\int_{-\infty}^{+\infty} \dd x\; p(x) e^{-i\omega x}
= \la e^{-i\omega x}\ra_{p}.
$$
Two random variables have the same characteristic function if and only
if they have the same PDF. Since the PDF of the sum of two independent
random variables $x_1$ and $x_2$ is the convolution product $p_1\star p_2$
of their PDF, the characteristic function of the sum is the product
$\chi_{1} \chi_{2}$ of the characteristic functions of $x_1$ and $x_2$.
This property makes the characteristic function a very useful
tool.\cite{Feller}

Once these mathematical objects are defined, let us ask the following question.
Consider a sequence of $N$ random variables $\{x_{k} \}_{k=1\cdots N}$,
distributed according to a joint probability distribution
$J_N(x_1,\ldots,x_N)$. Let us then define another random variable, say
$y_{N}=\phi(x_{1},\ldots,x_{N})$, where $\phi$ is a given arbitrary function.
The question we ask is whether it is
possible to find a sequence of numbers $\{(a_{N},b_{N}>0) \}$ such that
the distribution function of the random variable
$z_{N}=(y_{N}-a_{N})/b_{N}$ converges toward a PDF $p_{\infty}(z)$,
where $p_{\infty}(z)$ is a proper PDF (not concentrated at one point)
when $N$ goes to infinity.
In such case, $p_{\infty}(z)$ is called the asymptotic, or limit, distribution
of $z$, and by extension of $y$.
The joint probability distribution $J_N(x_1,\ldots,x_N)$ is said to
belong to the domain of attraction $\mathcal{D}(p_{\infty})$ of $p_{\infty}$.
In the case of i.i.d.~random variables, for which the joint
distribution factorizes as $J_N(x_1,\ldots,x_N)=\prod_{i=1}^N p(x_i)$,
one says that the marginal distribution $p(x)$ belongs to the domain
of attraction $\mathcal{D}(p_{\infty})$.

For a generic function $\phi(x_{1},\ldots,x_{N})$,
it is extremely difficult to identify the asymptotic distribution
(when it exists) and the corresponding domain of attraction.
However, two particular examples have been
extensively studied by mathematicians: the case of sums of random
variables,\cite{Kolmogorov,Levy,Feller2,Petrov75,Petrov95}
$\phi(x_{1},\cdots,x_{N})=\sum_{i=1}^Nx_{i}$, and the
case of extreme values, $\phi(x_{1},\cdots,x_{N}) =
\max(x_1,\ldots,x_N)$.\cite{Gnedenko,Galambos,Gumbel,Embrechts97,Reiss01,Falk}
As already pointed out in the introduction, we shall focus
in this article on these two specific cases, and see that
under additional assumptions, mathematics provides some
very interesting results.

As a side remark, let us note that
words like ``domain of attraction'' and ``limit function'' might ring a
bell to the physicist: it is indeed reminiscent of the renormalization
group language. This is actually not a mere coincidence, as pointed out
for instance by Jona-Lasinio.\cite{Jona01}
One can actually see the renormalization
group procedure as a way to compute limit distributions in physical
situations where standard mathematical results may not be applicable.
We shall come back to this point in Sect.~\ref{sect-RG}.

\section{Asymptotic distributions of sums and extremes of i.i.d.~random
variables}
\label{sect-convergence}

\subsection{Statistics of random sums and Gaussian distribution}

\subsubsection{Standard Central Limit Theorem for i.i.d.~random variables}
\label{sect-stdCLT}

Let us consider a set of random variables $\{x_i\}_{i=1,\ldots,N}$,
and let us denote their sum, which is also a random variable, as $y_N$:
\be
y_N = \phi(x_{1},\cdots,x_{N})= \sum_{i=1}^N x_i.
\ee
This new random variable could \textit{a priori} be described by an
arbitrary PDF. However, it turns out that the Gaussian (or normal)
distribution appears in such a large collection of phenomena that it
has been somehow considered as ``universal''.
This belief led Henry Poincar\'e to his famous comment:\cite{Poincare}
``\textit{All the world believes it firmly, since mathematicians
imagine that it is a fact of observation and the observers that it is a
mathematical theorem.}'' 
In this section we summarize the main standard
theorems, and comment on their applications in physics.

The case of sums of i.i.d.~random variables is clearly the simplest one.
In this case, there is only one single free distribution in the problem:
\be
J_N(x_1,\ldots,x_N)=\prod_{i =1}^N p(x_i),
\ee
and it is possible to establish the following standard
version of the Central Limit Theorem.\cite{Kolmogorov,Petrov75}

\begin{theorem}\label{CLT1}
Let $\{x_i \}_{i=1,\ldots,N}$ be a sequence of $N$ random variables,
independent and identically distributed, with a finite mean value $m$,
and a finite variance $\sigma$. Let us define $y_N = \sum_{i=1}^N x_i$,
$a_{N}=Nm$, and $b_{N}=\sigma \sqrt{N}$. Then the random variable
\be
z_N=\frac{y_N-a_{N}}{b_{N}}.
\ee
converges in law when $N\rightarrow \infty$ to the normal distribution
$\mathcal{N}_{0,1}$.
\end{theorem}
The normal distribution $\mathcal{N}_{0,1}$ is the Gaussian
distribution with zero mean and unit variance
\be
\mathcal{N}_{0,1}(z) = \frac{1}{\sqrt{2\pi}} \, e^{-z^2/2}.
\ee

Accordingly, one sees that three essential hypotheses are needed in order
to apply the Central Limit Theorem (at least in its standard form):
\begin{itemize}
\item the variables $x_i$ are independent.
\item the variables $x_i$ are identically distributed.
\item the second moment of $x_i$ is finite (the finiteness of the mean value
and of the variance then follows).
\end{itemize}

When at least one of these conditions is not fulfilled, the standard form
of the Central Limit Theorem is no longer valid. We see that given the
fact that the variables are i.i.d., this theorem applies to a large number of
probability distributions, as the only requirement is the existence
(i.e., finiteness) of the second moment.
In other words one could say that the basin of attraction
of the normal distribution is `large'.

It is possible to extend slightly the previous theorem to include some
distributions without second moment. The CLT for i.i.d.~random variables
in its general form is given below.\cite{Kolmogorov}

\begin{theorem} \label{CLT2}
Let $\{x_i \}_{i=1..N}$ be a sequence of $N$ random variables, independent
and identically distributed according to the probability distribution $p(x)$.
Then $p(x)$ belongs to the attraction basin of the normal law if and only if 
\bea \label{condCLT}
\lim_{X\rightarrow \infty} X^2\frac{\int_{|x|>X}\dd x\; p(x)}{\int_{|x|<X}
\dd x\; x^2p(x)}=0.
\eea
\end{theorem}

As an illustration of this extension, let us consider the case of a power-law
probability density $p(x)$:
\be
p(x)= \frac{\alpha\, x_0^{\alpha}}{x^{1+\alpha}} \Theta(x-x_0), \qquad (x_0>0).
\ee
If $\alpha>2$, the second moment of the distribution is finite,
and Theorem~\ref{CLT1} applies.
In constrast, if $\alpha \le 2$, the second moment is infinite,
and Theorem~\ref{CLT1} no longer holds.
However, for $\alpha=2$ we have
\be
X^2\frac{\int_{|x|>X}\dd x\; p(x)}{\int_{|x|<X}
\dd x\; x^2p(x)}=\frac{1}{2\ln(X/x_0)}
\longrightarrow 0,\; X\rightarrow \infty.
\ee
Theorem \ref{CLT2} then shows that this distribution is actually in the
attraction basin of the Normal law. Moreover similar calculations for
$\alpha<2$ show that 
\be
X^2\frac{\int_{|x|>X}\dd x\; p(x)}{\int_{|x|<X}
\dd x\; x^2p(x)}=\frac{2-\alpha}{\alpha} \frac{X^{2-\alpha}}{X^{2-\alpha}-
x_0^{2-\alpha}} \longrightarrow \frac{2-\alpha}{\alpha},\; X\rightarrow \infty.
\ee
These argument can be extended in a straightforward way to distributions
$p(x)$ with a power-law tail, $p(x) \sim x^{-1-\alpha}$, $x\rightarrow\infty$.
Therefore probability densities with power-law tails with $\alpha <2$
do not belong to the attractive basin of the Gaussian distribution.

Before discussing extreme value statistics, as well as several
extensions of the Central Limit Theorem, we would now like to briefly comment
on the usefulness of the CLT in physics.

\subsubsection{Examples of applications of the Central Limit Theorem
in physics}

The Central Limit Theorem is one of the cornerstones of statistical physics,
giving general results about fluctuations of global quantities
in systems where correlations among the different degrees of freedom
can be neglected.
As an illustration, let us consider a generic system with degrees of
freedom $\{q_i\}$, $i=1,\ldots, N$, described by a Hamiltonian which
can be written as the sum of one-body Hamiltonians:
\be
\mathcal{H}(\{q_i\}) = \sum_{i=1}^N \mathcal{H}_1(q_i)
\ee
where $\mathcal{H}_1(q_i)$ only depends on the single variable $q_i$
($q_i$ may be a real variable or a vector depending on the system considered).
At statistical equilibrium in the canonical ensemble, the joint distribution
of the variables $\{q_i\}$ is given by
\be \label{J-hamilt-indep}
J(\{q_i\}) = \frac{1}{\mathcal{Z}_1^N}\, \exp\left(-\beta \sum_{i=1}^N 
\mathcal{H}_1(q_i)\right)
= \prod_{i=1}^N \frac{1}{\mathcal{Z}_1} \, e^{-\beta \mathcal{H}_1(q_i)}.
\ee
The partition function factorizes as a product of $N$
one-body partition functions $\mathcal{Z}_1$, given by:
\be
\mathcal{Z}_1 = \int_{-\infty}^{\infty} \dd q \, e^{-\beta \mathcal{H}_1(q)}
\ee
Accordingly, the different degrees of freedom are i.i.d.~random variables.
In many cases, the variance
of $q_i$ is finite, so that the CLT can be applied to the sum
$S=\sum_{i=1}^N q_i$. Once suitably rescaled, this sum thus has a Gaussian
statistics in the large $N$ limit, independently of the specific form
of the one-body Hamiltonian (as long as the variance of $q_i$ remains finite).
Consider for instance a paramagnet consisting of an assembly of $N$ Ising
spins $s_i=\pm 1$, and with a Hamiltonian
\be
\mathcal{H}_{\mathrm{para}}(\{s_i\}) = -\sum_{i=1}^N h s_i
\ee
where $h$ is the external magnetic field. 
The variance of $s_i$ is obviously finite since $s_i$ is bounded.
From the CLT,
the total magnetization $M=\sum_{i=1}^N s_i$ is distributed according
to a Gaussian law in the thermodynamic limit $N \rightarrow \infty$.

Coming back to the general Hamiltonian $\mathcal{H}(\{q_i\})$,
another issue of physical interest is to determine the fluctuations
of the total energy of the system, that is of the value of
$\mathcal{H}(\{q_i\})$.
One then needs to make a change of variables, introducing
$\ve_i = \mathcal{H}_1(q_i)$, in order to rewrite the total energy as the sum
$E=\sum_{i=1}^N \ve_i$. From Eq.~(\ref{J-hamilt-indep}),
the variables $\ve_i$ are also i.i.d.~random variables, with distribution
\be
p(\ve_i) = \frac{1}{\mathcal{Z}_1}\, e^{-\beta\ve_i} |\mathcal{J}(\ve_i)|
\ee
where $\mathcal{J}(\ve_i)$ is the Jacobian of the transformation.
If the variance of $\ve_i$ is finite, then the CLT can be applied to
the total energy $E$ which then follows a Gaussian statistics.
As a simple example, let us consider a classical gas of independent particles.
In the absence of external field, the Hamiltonian reduces to the kinetic
energy:
\be
\mathcal{H}_{\mathrm{gas}}(\{\mathbf{p}_i\}) =
\sum_{i=1}^N \frac{\mathbf{p}_i^2}{2m}
\ee
The distribution of kinetic energy per particle $\ve_i=\mathbf{p}_i^2/2m$
is then, in $D$ dimensions,
\be
p(\ve_i) = \frac{\beta^{\frac{D}{2}}}{\Gamma\left(\frac{D}{2}\right)}
\, \ve_i^{\frac{D}{2}-1} e^{-\beta \ve_i},
\ee
where $\Gamma(t)=\int_0^{\infty} \dd u\, u^{t-1}e^{-u}$ is the Euler Gamma
function.
The second moment is finite, so that from the CLT, the total energy
follows a Gaussian statistics.

As we have seen, a large class of observables obey a Gaussian statistics
in equilibrium systems, provided that the energy of the coupling terms
in the Hamiltonian can be neglected.
When this is not the case, it is often possible at a heuristic level
to split the system into essentially independent ``blocks'' that play the
same role as the independent degrees of freedom in the above description.
Hence, the CLT is often effectively valid in physical systems, although
the strict hypotheses underlying it may not be fulfilled.
These issues are discussed in more details in Sect.~\ref{sect-non-iid}.

Note also that many other examples of the application of the CLT in statistical
physics could be given, where the variables to be summed are not
described by a Hamiltonian, specifically when considering out-of-equilibrum
systems. This is the case for instance for the (algebraic)
distance travelled by a one-dimensional random walk. A possible physical
realization is to follow the coordinate along a given axis of a Brownian
particle in a fluid. If the steps of the walk (i.e., the distance between
two successive collisions) are independent random variables
with finite second moment, then
the coordinate of the particle at a given large time is distributed
according to a Gaussian law.

\subsection{Extreme value statistics of i.i.d.~random variables}

\subsubsection{Universality classes and asymptotic distributions}
\label{iidext1}

Another example of convergence theorem is given by the mathematical
theory of extreme value statistics, that is the maximum or minimum value
in a set of random variables. Following Sect.~\ref{sect-asympt-dist}
this situation
corresponds to the function $\phi(x_1,\ldots,x_N)=\max(x_1,\ldots,x_N)$.
Extreme value statistics has found applications in many different fields,
like physics of disordered systems,\cite{BouchMez97}
chemical fracture,\cite{Baldassarri02} hydrology,\cite{Katz02}
seismology,\cite{Sornette96} or finance,\cite{Longin00,Potters}
to quote only some of them.

For the sake of simplicity we shall restrict ourselves
to the case of maximal values, but equivalent results are available for
minimal values, since a minimum can be converted into a maximum by simply
changing the sign of the random variable. The problem is the following.
Let $\{x_i\}_{i=1\dots N}$
be a set of $N$ i.i.d.~random variables, whose common cumulative distribution
is $F(x)$ (i.e., the probability that the random variable is less than the
given value $x$). From each realization of the set, one can define
a new random variable 
\be
z_N = \max(x_1,\ldots,x_N).
\ee
As by definition $x_i \le z_N$ for all $i$, the probability
$F_N(z)$ that the maximum is less than a value $z$ (that is, the cumulative
distribution of the maximum) is simply, from the i.i.d.~property,
\begin{eqnarray} \nonumber
F_N^{\mathrm{max}}(z) &\equiv& \mathrm{Prob}(\max(x_1,\ldots,x_N)<z)\\
&=& \mathrm{Prob}(\forall i, x_i<z),
\end{eqnarray}
so that one has
\be \label{eqext1}
F_N^{\mathrm{max}}(z)=F(z)^N.
\ee
A natural question is to try to determine the asymptotic cumulative
distribution of this maximum value $z_N$ (if it exists), with the hope
that it does not depend on all the
details of the original cumulative distribution $F$.
More precisely, one should wonder whether it would be possible to
find a sequence $\{a_N,b_N\}$ such that 
\be
\lim_{N \rightarrow \infty} F_N^{\mathrm{max}}(a_N + b_N x)= H(x),
\ee
where $H$ is a non-degenerate cumulative probability function to be
determined as well. We then have the following results.\cite{Galambos,Gumbel}

\begin{theorem}\label{th-EVS1}
The asymptotic cumulative distribution $H(x)$ of a set of i.i.d.~random
variables $\{x_i\}_{i=1,\ldots,N}$ is necessarily (up to a shift and a
dilatation of the variable $x$)
of the form $H_{f,\mu}(x)$, $H_{w,\mu}(x)$ or $H_g(x)$ with
\begin{itemize}
\item $H_{f,\mu}(x)=\exp\left(-x^{-\mu} \right)$
for $x>0$ and $0$ for $x\le0$ (Fr\'echet distribution); 
\item $H_{w,\mu}(x)= \exp \left( -(-x)^\mu \right)$
for $x<0$ and $1$ for $x \ge 1$ (Weibull distribution);
\item $H_g(x) = \exp \left(-e^{-x} \right)$ (Fisher-Tippett-Gumbel
or Gumbel distribution).
\end{itemize} 
\end{theorem}

In the previous expressions, $\mu>0$ is a parameter, called the extreme
value index, depending on the parent cumulative probability
distribution $F(x)$.
Note that a more compact parameterization has been proposed by
von Mises\cite{Mises}. Up to a dilatation and a shift of the variable $x$,
the asymptotic extreme value distributions $H_{f,\mu}(x)$, $H_{w,\mu}(x)$
and $H_g(x)$ can be reformulated as
$H_\gamma(x)$, with $a>0$ and $b$ real, where
\be \label{eq-vonMises}
H_\gamma(x)=\exp\left( -(1+\gamma x)^{-1/\gamma}\right),\; 1+\gamma x >0.
\ee
The parameter $\gamma$ is real and the value $\gamma=0$ in the right-hand
side should be interpreted as $\exp \left(-e^{-x}\right)$, then corresponding
to the Gumbel distribution.
The case $\gamma >0$ corresponds to the Fr\'echet distribution, while
$\gamma <0$ stands for the Weibull case.

Hence the parameter $\mu$ in Theorem~\ref{th-EVS1} is related to the parameter
$\gamma$ in Eq.~(\ref{eq-vonMises}) through $\mu=1/\gamma$ if $\gamma>0$
(Fr\'echet distribution) and $\mu=-1/\gamma$ if $\gamma<0$ (Weibull
distribution).
Now we have to specify the attraction basin $\mathcal{D}(H_{\gamma})$
for each of the previous distributions. This is the purpose of the following
theorem.\cite{Falk}

\begin{theorem} \label{th-attrac-basin}
Let $F(x)$ and $H(x)$ be non-degenerate cumulative distribution functions,
such that for some constants $a_N>0$ and $b_N$ real
$$
\lim_{N \rightarrow \infty} F(a_N+b_N x)^N = H(x).
$$
Then $H$ is up to a location and scale shift an extreme value distribution
$H_\gamma$, and $F$ belongs to the attraction basin $\mathcal{D}(H_\gamma)$
of the distribution $H_\gamma$. We also define $\tilde{F}(z)=1-F(z)$,
which satisfies $\lim_{z \rightarrow \infty}\tilde{F}(z)=0$.
Let $\omega_F=\sup \{ z : \tilde{F}(z)>0\}$. We then have:
\begin{itemize}

\item $F$ belongs to the attraction basin
$\mathcal{D}(H_{f,\mu})=\mathcal{D}(H_{\gamma=1/\mu})$ if, and
only if, $\omega_F=+\infty$ and for all $x>0$,
\be
\lim_{t \rightarrow \infty} \frac{\tilde{F}(tx)}{\tilde{F}(t)}= x^{-\mu},
\quad \mu>0 ;
\ee

\item $F$ belongs to the attraction basin
$\mathcal{D}(H_{w,\mu})=\mathcal{D}(H_{\gamma=-1/\mu})$ if,
and only if, $\omega_F$ is finite and for all $x>0$,
\be
\lim_{t \rightarrow \infty} \frac{\tilde{F}\left(\omega_F-\frac{1}{tx} \right)}
{\tilde{F}\left(\omega_F-\frac{1}{t} \right)}= x^{-\mu},\quad \mu>0 ;
\ee

\item $F$ belongs to the attraction basin
$\mathcal{D}(H_{g})=\mathcal{D}(H_{\gamma=0})$ if, and only if,
for some finite $\omega_0$ the integral
$\int_{\omega_0}^{\omega_F} \dd z\, \tilde{F}(z)$ is finite and for all $x>0$,
\be
\lim_{t \rightarrow \infty} \frac{\tilde{F}\left(t+xR(t)\right)}{\tilde{F}(t)}= e^{-x},
\ee
with $R(t)$ defined by
\be
R(t)=\frac{1}{\tilde{F}(t)}\int_t ^{\omega_F} \dd z\, \tilde{F}(z),
\ee

\end{itemize}
\end{theorem}

Note that it is also customary to speak about classes instead of basins
of attraction, in the context of extreme value statistics. For instance,
the exponential distribution is said to be in the Gumbel class.
Let us now try to give an intuitive interpretation of
Theorem~\ref{th-attrac-basin}.
Essentially, it means that when $\tilde{F}(z)$, namely the probability that the
random variable is greater than $z$, decays like a power law at large $z$,
$\tilde{F}(z)\sim z^{-\mu}$ with $\mu>0$, the limit distribution is the
Fr\'echet one with parameter $\mu$.
When $z$ is a bounded variable, that is $\tilde{F}(z)=0$ for $z>\omega_F$,
and if $\tilde{F}(z) \sim (\omega_F-z)^{\mu}$,
$z \rightarrow \omega_F^{-}$ with $\mu>0$, then the asymptotic
distribution is the Weibull one with parameter $\mu$.
Finally, if $z$ is unbounded, but if $\tilde{F}(z)$ decays faster than any
power law when $z \rightarrow \infty$, the asymptotic distribution is
the Gumbel one. Note however that the theorem also leads to the Gumbel
distribution in some cases of bounded variables, like for instance
$\tilde{F}(z)\sim \exp[-c(\omega_F-z)^{-\alpha}]$, $z\rightarrow \omega_F^{-}$,
where $\alpha$ and $c$ are positive constants.

As an illustration of how these limit laws can be derived,
let us consider the simple example of the exponential
distribution, for which the cumulative probability distribution is
$F(z) = [1-\exp(-\lambda z)]$ for $z>0$ and $F(z)=0$ for $z\le 0$.
In this case, the cumulative distribution $F_N^{\mathrm{max}}(z)$,
i.e., the probability
that the largest value in the set in $N$ variables drawn from $F$ is
smaller than $z$, reads
\be
F_N^{\mathrm{max}}(z) = F(z)^N = \left( 1-e^{-\lambda z} \right)^N
\ee
On general grounds, one wishes to find a rescaling $z=a_N+b_N x$ such that
the distribution of $x$ converges to a well-defined limit when $N$
goes to infinity. In the present case, it is rather easy to see that
a correct choice for the rescaling is $a_N=(\ln N)/\lambda$ and
$b_N=1/\lambda$, yielding
\be
F_N^{\mathrm{max}}(a_N+b_N x) = \left( 1-\frac{e^{-x}}{N} \right)^N
\rightarrow \exp\left( e^{-x}\right)
\ee
when $N \rightarrow \infty$.
Hence the distribution $F_N^{\mathrm{max}}(a_N+b_N x)$ readily converges
to the Gumbel distribution $H_g(x)$.
Clearly, the rigorous derivation of the asymptotic distributions
in the case of an arbitrary $F(z)$ is much more difficult.
Yet, the main ideas of the derivation are already sketched in this
simple example.

\subsubsection{Asymptotic distributions for the $k^{th}$ largest value}

Up to now we only considered the asymptotic distribution of the largest
value within a set of $N$ i.i.d.~random variables, in the limit $N \rightarrow \infty$.
A further natural question would then be to know how the $k^{\mathrm{th}}$
largest value in the random set is distributed. Note that here $k$ is
independent of $N$, the number of elements in the set; the case where $k=k(N)$,
sometimes referred to as order statistics, could also be studied,\cite{Gumbel}
but is outside the scope of the present review.
Let us consider the cumulative probability $F_{k,N}(x)$ of the
$k^{\mathrm{th}}$ largest value $X_{k,N}$ in a set of $N$ i.i.d.~random
variables.
Then relation~(\ref{eqext1}) can be generalized to obtain\cite{Galambos}
\be
F_{k,N}(x)=\mathrm{Prob}\left(X_{k,N} \le x \right)= \sum_{j=0}^{k-1}
\frac{N!}{j!(N-j)!} F(x)^{N-j} [1-F(x)]^{j},
\ee
Keeping fixed the index $k$ while
$N \rightarrow \infty$, it turns out that the asymptotic form of $F_{k,N}$
is related to the one of $F_{k=1,N}$. This is expressed by the
following theorem\cite{Galambos}

\begin{theorem} \label{kext}
For a sequence $a_N$ and $b_N>0$ of real numbers and for a fixed integer
$k>1$, $F_{k,N}(a_N+b_N x)$ converges when $N \rightarrow \infty$
toward a non-degenerate cumulative probability $H_k(x)$ if and only if
$F_{1,N}(a_N+b_N x)$ converges weakly to a non-degenerate cumulative
probability $H(x)$, being one of the possible functions given by
Theorem~\ref{th-EVS1}.
If $H_k(x)$ exists then it is related to $H(x)$ through
\be
H_k(x)=H(x) \sum_{j=0}^{k-1}\frac{1}{j!}\left(\ln \frac{1}{H(x)} \right)^j.
\ee
\end{theorem}

\begin{figure}[!htb]
\begin{center}
\subfigure[Linear scale]{\includegraphics[width=0.45\linewidth]{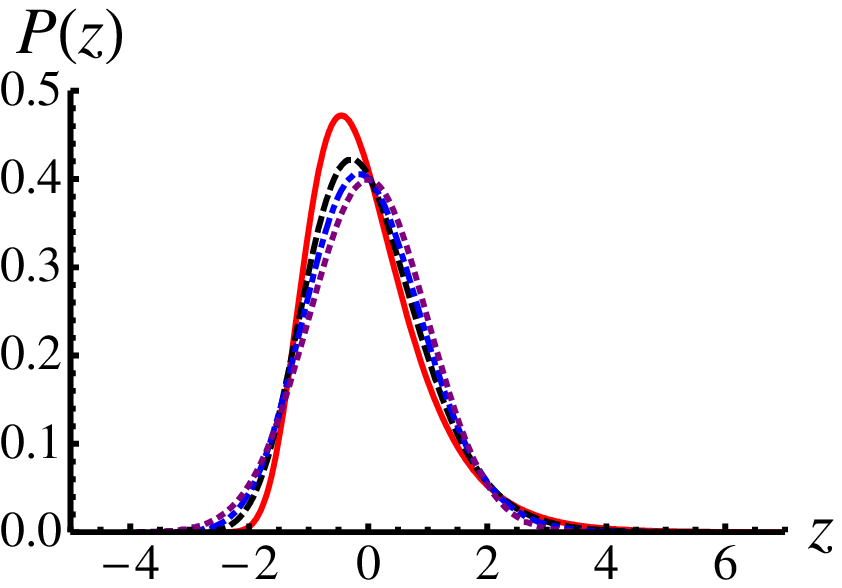}}
\subfigure[Semi-logarithmic scale]{\includegraphics[width=0.45\linewidth]{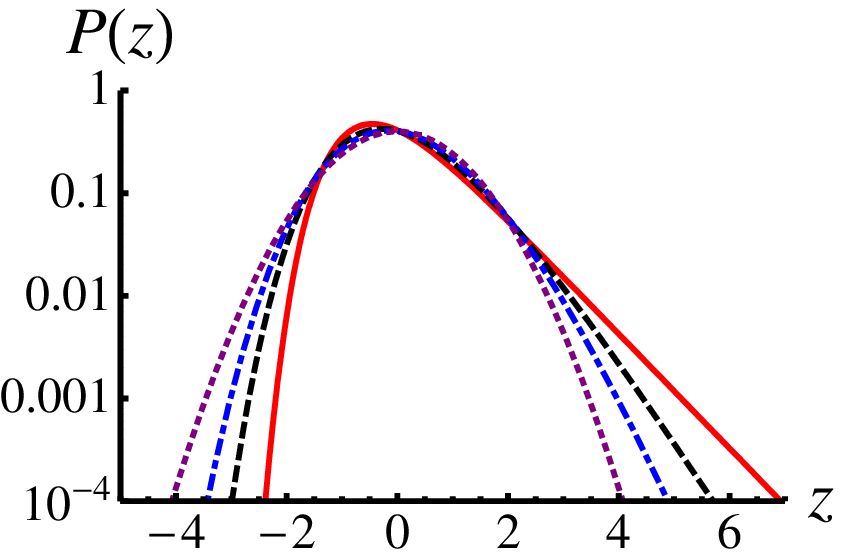}}
\caption{Gumbel distribution $g_k(x)$ for $k=1$ (plain), $3$ (dash) and $10$ (dot dash). The distributions are normalised to have a zero mean and a variance unity. Note the convergence to the Gaussian distribution (dot) when $k$ increases. }
\label{fig-gumbel}
\end{center}
\end{figure}

In terms of probability density functions $h_k(x)=dH_k/dx$, the relation
between the $k^{\mathrm{th}}$ largest and the largest value reads
\be
h_k(x)=\frac{h(x)}{(k-1)!} \left(\ln \frac{1}{H(x)} \right)^{k-1}.
\ee
For instance, for random variables belonging to the Gumbel class,
we have $H(x)=H_g(x)$ and
\be
h^{(k)}(x)=\frac{1}{(k-1)!}\exp \left(-kx-e^{-x} \right).
\ee
In order to compare experimental or numerical data with theoretical
predictions, it is convenient to make a slightly different choice
for the parameters $a_N$ and $b_N$, in such a way that some specific
moments of the distribution are normalized to zero or one.
This leads us to what we shall call in this
paper the ``standard forms'' of the generalized extreme value
distributions.
Namely, one finds for the generalized Gumbel distribution with parameter $k$
\bea \label{eqGk}
g_k(x)=\frac{k^k \theta_k}{\Gamma(k)} \exp \left[-k\theta_k(x+\nu_k)-
ke^{-\theta_k(x+\nu_k)}\right],\\
\nonumber \theta_k^2=\Psi'(k),\quad
\nu_k=\frac{1}{\theta_k}\left[\ln k - \Psi(k)\right],
\eea
where $\Psi(x)=\frac{\dd}{\dd x} \ln \Gamma(x)$ is the digamma function.
The generalized Gumbel distribution is illustrated on Fig.~\ref{fig-gumbel}
for different values of $k$.
Note that one can show that when $k$ goes to infinity, the distribution
$g_k(x)$ converges to the Gaussian distribution with zero mean and
unit variance, while it converges to an exponential distribution
for $k \rightarrow 0$.\cite{BCH08}

\begin{figure}[!htb]
\begin{center}
\subfigure[Linear scale]{\includegraphics[width=0.45\linewidth]{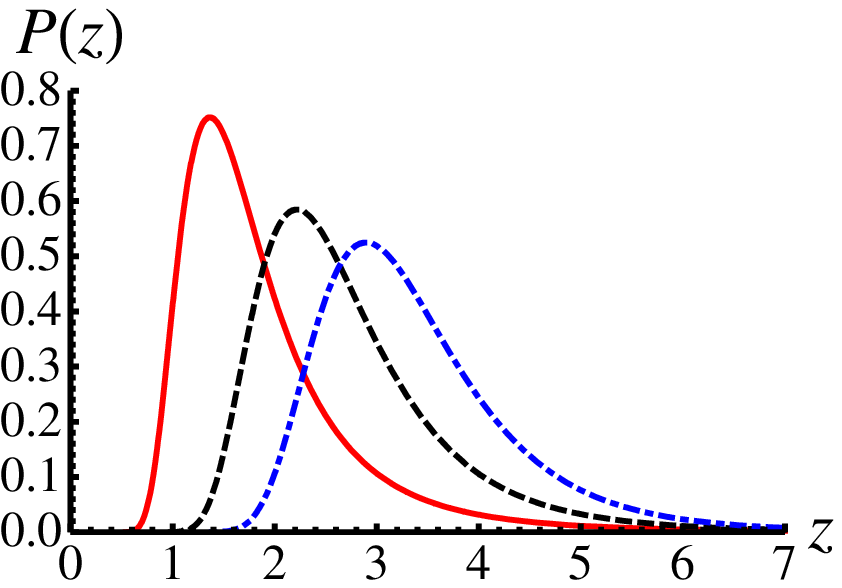}}
\subfigure[Semi-logarithmic scale]{\includegraphics[width=0.45\linewidth]{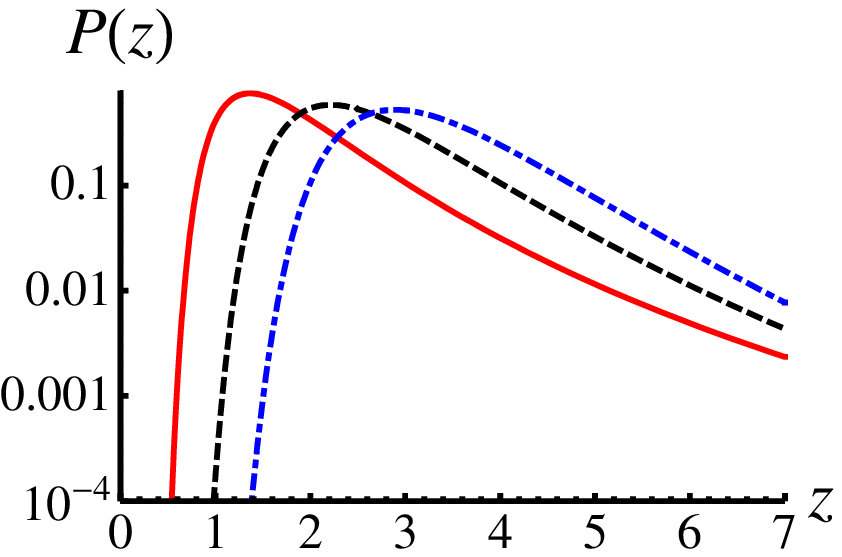}}
\caption{Fr\'echet distribution $f_{k,\mu}(x)$ for $k=2$, $3$ and $4$, with $\mu=2$. Distributions are normalised to have a variance unity.}
\label{fig-frechet}
\end{center}
\end{figure}

\begin{figure}[!htb]
\begin{center}
\subfigure[Linear scale]{\includegraphics[width=0.45\linewidth]{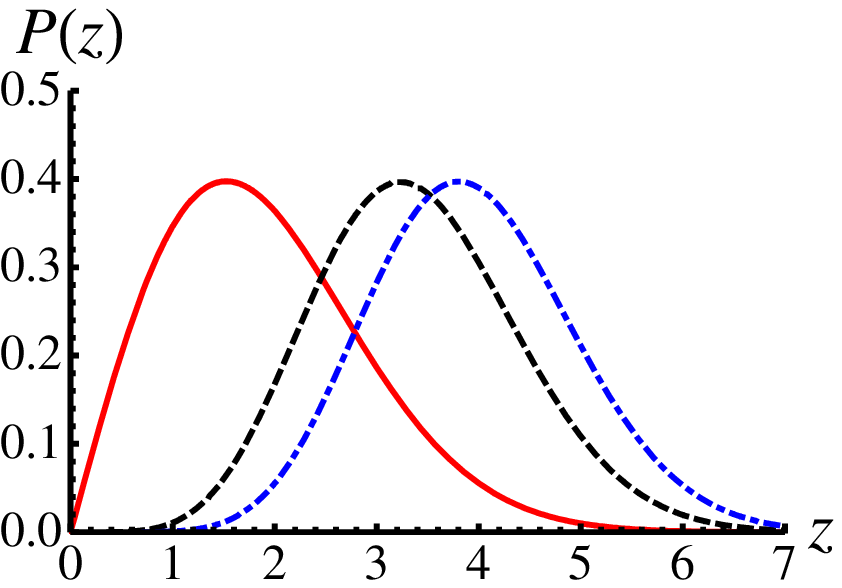}}
\subfigure[Semi-logarithmic scale]{\includegraphics[width=0.45\linewidth]{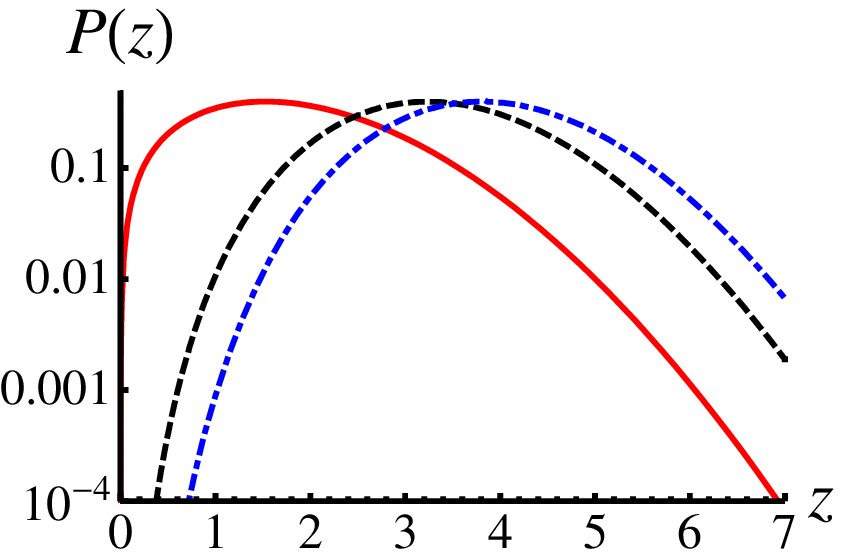}}
\caption{Weibull distribution $w_{k,\mu}(x)$ for $k=1$, $3$ and $4$, with $\mu=2$.Distributions are normalised to have a variance unity.}
\label{fig-weibull}
\end{center}
\end{figure}

The generalized Fr\'echet distribution is given for $x>0$ by
\bea \label{eqFk}
f_{k,\mu}(x)&=&\frac{\mu \lambda_f^k}{\Gamma(k)}\frac{1}{x^{1+k\mu}}\exp\left(
-\lambda_f x^{-\mu}\right),\\ \nonumber
\lambda_f&=&\Gamma(k)^\mu \left[\Gamma(k)\Gamma\left(k-\frac{2}{\mu}\right)
-\Gamma\left(k-\frac{1}{\mu}\right)^2\right]^{-\mu/2}.
\eea
and by $f_{k,\mu}(x)=0$ for $x \le 0$.
The generalized Weibull distribution reads for $x<0$
\bea \label{eqWk}
w_{k,\mu}(x)&=&\frac{\mu\lambda_w^k}{\Gamma(k)} (-x)^{\mu k -1}
\exp\left(-\lambda_w (-x)^\mu\right),\\ \nonumber 
\lambda_w&=&\Gamma(k)^{-\mu}\left[\Gamma(k)\Gamma\left(k+\frac{2}{\mu}\right)
-\Gamma\left(k+\frac{1}{\mu}\right)^2 \right]^{\mu/2}.
\eea
and $w_{k,\mu}(x)=0$ for $x \ge 0$.
Here the Fr\'echet and Weibull distributions are normalized such that
their variance is normalized to one. For the
Fr\'echet distribution, this is possible only when the first moment is
finite, that is for $k>1/\mu$.
The Fr\'echet and Weibull distributions are shown on Fig.~\ref{fig-frechet}
and Fig.~\ref{fig-weibull} respectively.

In the context of extreme value statistics, $\Gamma(k)=(k-1)!$
as the parameter $k$ is by definition an integer. 
Still, once written using Gamma functions, these distributions can formally
be extended to positive real values of $k$, although one then looses
a direct interpretation in terms of extreme value statistics.
We shall come back to the interpretation of these distributions with
noninteger values of $k$ in section \ref{generalization}. To conclude this
discussion let us mention that some results exist concerning the speed
of convergence towards asymptotic extreme value
distributions,\cite{Galambos,Falk,Racz07}
but such results go beyond the scope of the present review.

\subsection{Broad distributions: when extreme values change the statistics
of sums}
\label{sect-Levy-laws}

\subsubsection{A simple scaling argument on the largest term in the sum}

As mentioned in Section~\ref{sect-stdCLT}, the standard Central Limit
Theorem, associated to a Gaussian asymptotic distribution, breaks down
as soon as the second moment of the individual variables $x_i$ is
divergent, i.e., formally infinite (note however the slight extension
allowed by Theorem~\ref{CLT2}). 

Generically, distributions
$p(x)$ leading to such a divergence of $\la x_i^2 \ra$ are characterized
by a power-law decay at large argument (we assume here for simplicity that
$x>0$)
\be
p(x) \sim \frac{c}{x^{1+\alpha}}, \qquad x \rightarrow \infty\qquad (\alpha>0).
\ee
In this case, $x^q p(x)$ behaves at large $x$ as $1/x^{1+\alpha-q}$,
so that the $q^{\rm th}$ moment $\la x^q \ra$ converges only if $q<\alpha$.
Accordingly, the second moment $\la x^2 \ra$ is infinite as soon as
$\alpha \le 2$, and the first moment $\la x \ra$ is infinite
if $\alpha \le 1$.

This last case is indeed very instructive in order to get an intuitive
feeling of the somehow surprising behavior of sums of broadly
distributed variables.
As mentioned in the introduction, such sums are often dominated
by the largest terms. To see qualitatively how this happens, let us
estimate the largest value among the set $\{x_i\}$.
Introducing $\tilde{F}(z)\equiv \int_z^{\infty} p(x) \dd x$, one has
from Eq.~(\ref{eqext1})
\be
F_N^{\mathrm{max}}(z) \equiv 
\mathrm{Prob}(\max(x_1,\ldots,x_N)<z) = [1-\tilde{F}(z)]^N
\ee
For large $N$, relevant values of $z$ are also
typically large, so that $\tilde{F}(z)$ is small, and can be approximated
as $\tilde{F}(z) \approx c' z^{-\alpha}$, with $c'=c/\alpha$.
One then obtains the asymptotic large $N$ expression:
\be
F_N^{\mathrm{max}}(z) \approx e^{-N\tilde{F}(z)} \approx
e^{-c'Nz^{-\alpha}}.
\ee
Accordingly, the cumulative distribution $F_N^{\mathrm{max}}(z)$
can be written as a function
of the rescaled variables $z/N^{1/\alpha}$, which shows that typical
values of the maximum of the $x_i$'s are of the order of $N^{1/\alpha}$.

Coming back to the sum $\sum_{i=1}^N x_i$, one expects
from the laws of large numbers that the typical value of the sum is
proportional to $N$ at large $N$, namely
$\sum_{i=1}^N x_i \approx N \langle x \rangle$.
This should be true at least when $\langle x \rangle$ is finite.
Indeed, if $\alpha >1$, the largest term in the sum
is of the order of $N^{1/\alpha}$, and becomes negligible with respect
to $N$ at large $N$. Hence the largest term does not dominate the sum.
In contrast, if  $\alpha <1$, $N^{1/\alpha} \gg N$ at large $N$, so that
the sum can no longer be proportional to $N$ (which is consistent with
the divergence of $\langle x \rangle$). In this case, the largest
term dominates the sum, and the latter scales, similarly to the largest
term, as $N^{1/\alpha}$.
Note however that the total contribution of the other terms in the sum does
not become negligible even when $N \rightarrow \infty$.
Only the scaling
behavior of the sum with $N$ is the same as that of the largest terms,
but the total contribution of the other terms is itself of the order
of $N^{1/\alpha}$.
This explains why, although sums of broadly distributed random
variables are dominated by extreme values (the largest terms),
the distribution of such sums does not belong to the
classes of asymptotic extreme value distributions. Note however that
both the L\'evy-stable distributions (see Sect.~\ref{sect-GCLT} below)
and the Fr\'echet distribution share a power-law tail
with the same exponent.

\subsubsection{Generalized Central Limit Theorem and L\'evy-stable laws}
\label{sect-GCLT}

In the case when the second moment $\la x_i^2 \ra$ is infinite, the standard
Central Limit Theorem no longer applies (apart from the slight extension
given in Theorem~\ref{CLT2}). Yet, it is still possible
to find rescaling parameters $a_N$ and $b_N$ such that the distribution
of the rescaled random variable $z_N$,
\be
z_N = \frac{1}{b_N} \left(\sum_{i=1}^N x_i - a_N\right)
\ee
converges to a limit distribution when $N \rightarrow \infty$, which belongs to
a family of distributions called L\'evy-stable laws.
These stable distributions depend
on two shape parameters $\alpha$ and $\beta$,
with $0<\alpha \le 2$ and $-1 \le \beta \le 1$, and two scales parameters.
This means that keeping the shape of the distribution constant, one
can always build another stable distribution through a translation and
dilatation of the variable, as in the case of extreme value statistics.
Since these scale parameters can be eliminated through a redefinition of
$a_N$ and $b_N$, we shall not make explicit reference to them in the
following.
Hence, we shall make a standard choice of scale parameters,
and simply denote the L\'evy-stable laws as $L(z;\alpha,\beta)$.
The following generalized central limit theorem for sums of broadly
distributed variables then holds.\cite{Kolmogorov,Levy,Feller2,BG}

\begin{theorem} \label{GCLT} Let $\{x_k\}_{k=1,\ldots,N}$
be a sequence of i.i.d.~random
variables with cumulative probability distribution $F(x)$.
We define the random variable $y_N=\sum_{k=1}^N x_k$.
Let $\alpha$ and $\beta$ be two real numbers such that $0<\alpha\le 2$
and $-1\le \beta \le 1$. Then $F(x)$ belongs to
the attraction basin of the L\'evy distribution $L(z;\alpha,\beta)$
defined by its characteristic function
\begin{eqnarray}
\hat{L}(k;\alpha,\beta) &=& \exp[-|k|^{\alpha}(1+i\beta \mathrm{sgn}(k)
\varphi(k,\alpha))],\label{levyTF}\\
\varphi(k,\alpha) &=& \tan \frac{\pi\alpha}{2} \qquad \quad \mathrm{if} 
\qquad \alpha \ne 1,\\
\varphi(k,\alpha) &=& \frac{2}{\pi}\ln|k| \qquad \quad \mathrm{if} 
\qquad \alpha = 1,
\end{eqnarray}
if the following conditions hold:
\begin{eqnarray} \label{eq-beta}
&&\lim_{x \rightarrow \infty} \frac{F(-x)}{1-F(x)}=\frac{1-\beta}{1+\beta},\\
&&\lim_{x \rightarrow \infty} \frac{1-F(x)+F(-x)}{1-F(rx)+F(-rx)}
=r ^{\alpha},\: \forall r>0.
\label{eq-alpha}
\end{eqnarray}
\end{theorem}
The scale parameters are chosen such that
the Fourier transform $\hat{L}(k;\alpha,\beta)$ of $L(z;\alpha,\beta)$
has the simple form (\ref{levyTF}).
Apart from a few specific values of $\alpha$ and $\beta$, no explicit
expression for $L(z;\alpha,\beta)$ is known in general, but
asymptotic expressions (for instance, at large $|z|$)
can be derived from the Fourier transform.
For $\alpha=2$, $\varphi(k,2)=0$, and
\be
\hat{L}(k;2,\beta) = \exp(-k^2)
\ee
for all values of $\beta$, so that $L(z;2,\beta)$ is simply a Gaussian
distribution.
For $\alpha<2$, $L(z;\alpha,\beta)$ depends on $\beta$,
which characterizes the asymmetry of the distribution; the case $\beta=0$
then corresponds to a symmetric limit distribution.
Another case where the distribution is known explicitly is when
$\alpha=1$ and $\beta=0$, corresponding to the Cauchy distribution:
\be
L(z;1,0) = \frac{1}{\pi(1+z^2)}.
\ee
\begin{figure}[t]
\begin{center}
\subfigure[\label{Levy1lin}Linear scale.]{\includegraphics[width=0.45\linewidth]{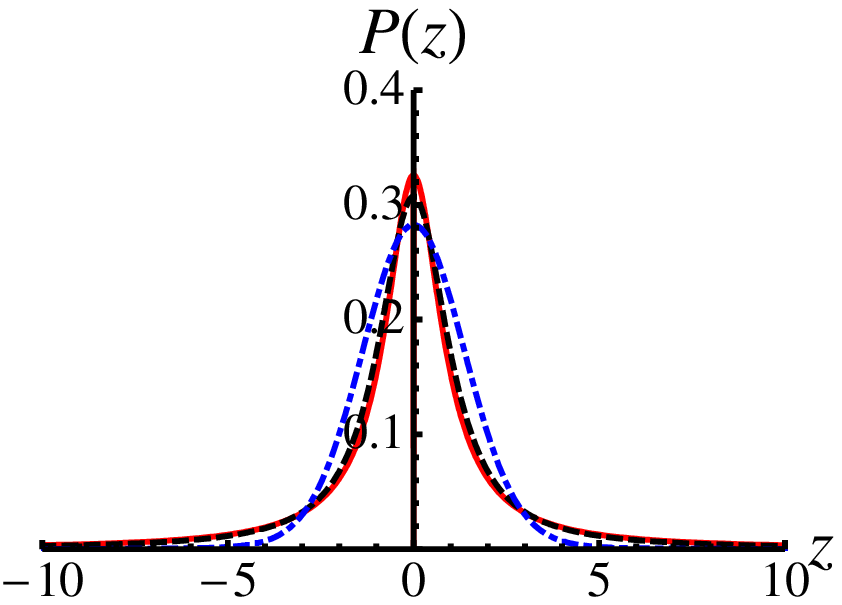}}
\subfigure[\label{Levy1log}Logarithmic scale.]{\includegraphics[width=0.45\linewidth]{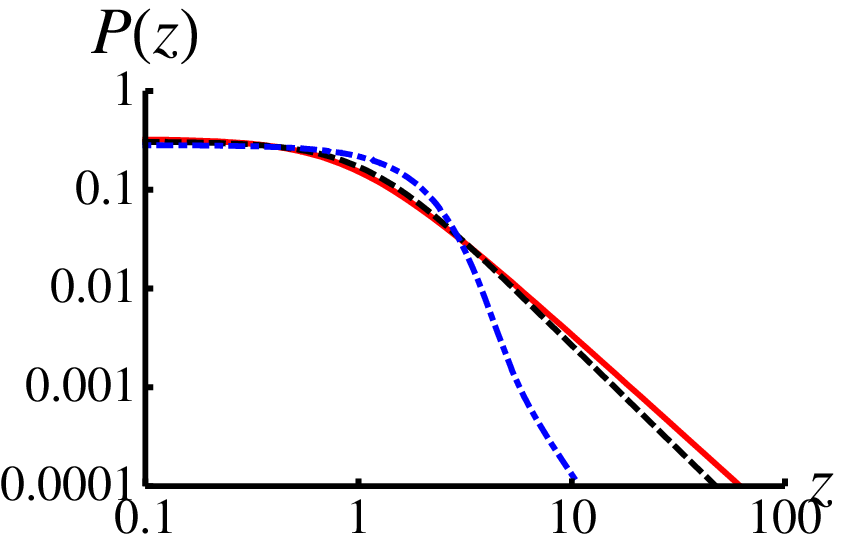}}
\caption{Examples of L\'evy distributions, for $\beta=0$ and $\alpha=0.95$ (plain), $\alpha=1.1$ (dash) and $\alpha=1.9$ (dot dash).}
\end{center}
\end{figure}

\begin{figure}[t]
\begin{center}
\subfigure[\label{Levy2lin} Linear scale.]{\includegraphics[width=0.45\linewidth]{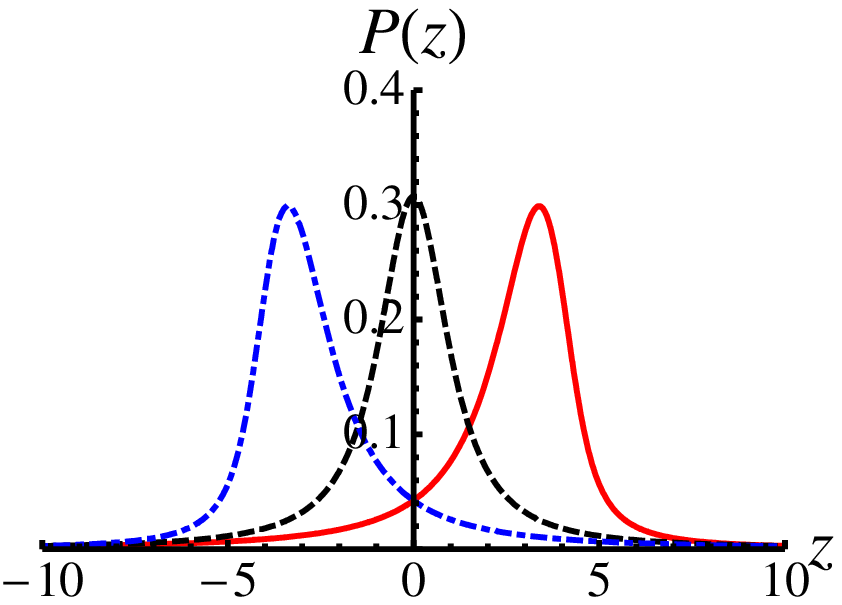}}
\subfigure[\label{Levy2log}Logarithmic scale.]{\includegraphics[width=0.45\linewidth]{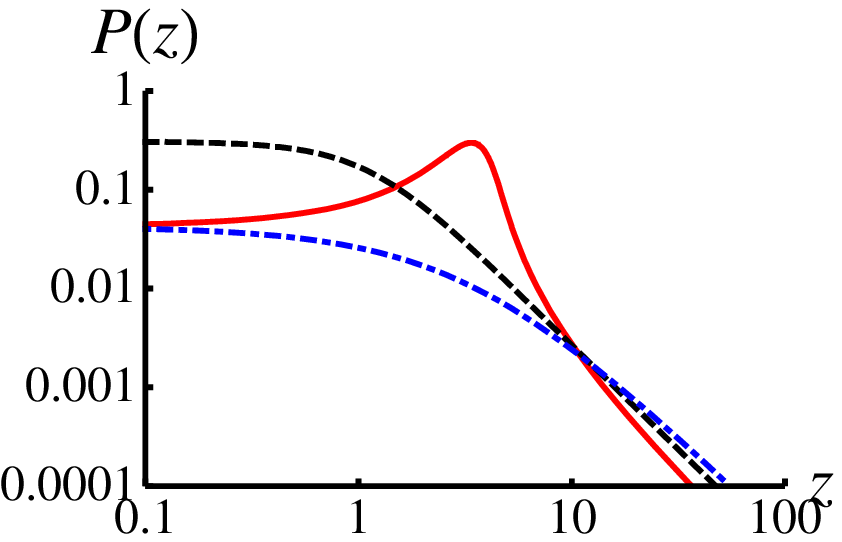}}
\caption{Examples of L\'evy distributions, for $\alpha=1.1$ and $\beta=-0.5$ (plain), $\beta=0$ (dash) and $\beta=0.5$ (dot dash).
Distributions have been centered such that $\langle z \rangle=0$.}
\end{center}
\end{figure}

In practice, the parameters $\alpha$ and $\beta$ can be obtained from
the distribution $p(x)=dF/dx$ in the following way.
If $p(x)$ behaves asymptotically as
\begin{eqnarray}
p(x) &\sim& \frac{c_{+}}{x^{1+\alpha_{+}}}, \qquad x \rightarrow +\infty,\\
p(x) &\sim& \frac{c_{-}}{|x|^{1+\alpha_{-}}}, \qquad x \rightarrow -\infty,
\end{eqnarray}
with $0<\alpha_{+},\alpha_{-}\le 2$,
then from Eqs.~(\ref{eq-beta}) and (\ref{eq-alpha}), $\alpha$ and $\beta$
are given by
\begin{eqnarray}
\alpha &=& \min(\alpha_{+},\alpha_{-}),\\
\beta &=& \frac{c_{+}-c_{-}}{c_{+}+c_{-}} \qquad \mathrm{if}
\quad \alpha_{+} = \alpha_{-},
\end{eqnarray}
whereas $\beta=1$ if $\alpha_{+} < \alpha_{-}$ and $\beta=-1$ if
$\alpha_{+} > \alpha_{-}$. 
When $\beta=1$, the distribution $L(z;\alpha,\beta)$ vanishes for $z<0$;
similarly, it vanishes for $z>0$ if $\beta=-1$.
This means that for instance a broad negative
tail with exponent $\alpha_{-}$ actually ``disappears'' from the limit
distribution if the positive tail is even broader, that is if
$\alpha_{+} < \alpha_{-}$.
Note that when $L(z;\alpha,\beta)$ is nonzero on a given domain
($z<0$ or $z>0$), its corresponding tail has the same exponent
as the original distribution $p(x)$ on the same domain.
For completeness, let us also mention the behavior with $N$ of the
rescaling coefficients $a_N$ and $b_N$.
The coefficient $b_N$ is of the form $b_N=b_0 N^{1/\alpha}$,
while $a_N=N\langle x\rangle+a_0$ if $1<\alpha<2$ and $a_N=0$
if $0<\alpha<1$, where $a_0$ and $b_0$ are some constants.
The case $\alpha=1$ involves logarithmic corrections.

\subsubsection{Example of application: laser cooling of atoms}

Although many systems obey the standard CLT, numerous different examples
of broad distribution effects in physical systems have been reported in the
last decades. To give only a few examples, these range from
glassy dynamics,\cite{Shl88,Bouchaud92,Monthus96,Bertin02}
anomalous diffusion in disordered
media,\cite{BG,Rinn00,Bertin03,Bertin03b,Monthus03}
diffusion of spectral lines in disordered solids,\cite{ZuK94,Barkai}
and random walks in solutions of micelles,\cite{OBL90} to
laser cooling of atomic gases,\cite{BBE94,BBA02}
laser trapped ions,\cite{MEZ96,KSW97}
fluorescence of single nanocrystals,\cite{Brokmann}
or chaotic transport\cite{SZK93,SWS93} and turbulent flows\cite{Min96}.

As an illustration,
let us consider the case of laser cooling experiments, which consist
in reducing the dispersion of momenta of atoms through interaction
of the atoms with photons.
One of such cooling protocols is called subrecoil laser cooling,
and it consists in reducing the momentum spread of the atoms
to a value much smaller than the typical momentum $\hbar k$ of the
photons.\cite{BBE94,BBA02,Bardou05}
Denoting by $q$ the magnitude of the atomic momentum $\mathbf{q}$,
the atom-photon
interaction process is such that the scattering rate of the photon $R(q)$
vanishes for $q=0$. In most cases, $R(q)$ behaves in the vicinity of $q=0$
as a power law of $q$ with an even exponent: $R(q) \sim q^{\beta}$,
$q \rightarrow 0$, with $\beta=2$, $4$, or $6$, depending
on the specific atom-photon interaction process.
This means that the average time spent at momentum $q$ between two photon
scatterings, to be denoted as the lifetime $\tau(q)$, behaves at
small $q$ as a diverging power law of $q$:
\be
\tau(q) = \tau_0 \left( \frac{q_m}{q} \right)^{\beta}
\ee
where $\tau_0$ and $q_m$ are time and momentum scales respectively
(note that $q_m \ll \hbar k$).
After a photon scattering, the atom typically has a momentum
$q \approx \hbar k$, so that the scattering rate becomes high. Then
successive scatterings will rapidly make the atomic momentum come back
to a value $q < q_m$, where it will stay again for a long time.
Neglecting the time spent in the region $q>q_m$, and assuming that
the values of the momentum $\mathbf{q}$ are uniformly distributed
in the $D$-dimensional sphere of radius $q_m$, the distribution of $q$
reads:
\be
\rho(q) = D \frac{q^{D-1}}{q_m^D}, \qquad 0<q<q_m.
\ee
Let us now determine the distribution $\psi(\tau)$ of the lifetime $\tau$.
From the relation $\psi(\tau)|d\tau|=\rho(q)|dq|$, one finds
\be
\psi(\tau) = \frac{c}{\tau^{1+\alpha}}, \qquad \tau > \tau_{\mathrm{min}},
\ee
where $c$ and $\tau_{\mathrm{min}}$ are parameters that can be expressed
as a function of $\tau_0$, $p_m$ and $D$. The exponent $\alpha$ is given
by $\alpha = D/\beta$. Note that $\tau$ is the average time spent at
momentum $q$, and that the actual time $\tau_s$ spent at momentum
$q$ fluctuates according to an exponential distribution
of mean $\tau$. Despite these fluctuations, the distribution
of $\tau_s$ averaged over $q$ behaves similarly to $\psi(\tau)$
at large $\tau$, so that we shall not distinguish between the
two notions in what follows.

The time $T_N$ of the $N^{\mathrm{th}}$ scattering can be expressed as
$T_N = \sum_{i=1}^N \tau_i$. If $\alpha<2$, the fluctuations of $T_N$
at large $N$ are no longer Gaussian, but distributed according to the
L\'evy-stable distribution of index $\alpha$, and asymmetry parameter
$\beta=1$. If $\alpha<1$, the
typical value of $T_N$ is no longer proportional to $N$, but rather
to $N^{1/\alpha}$, since the formal average value of $\tau$ is infinite.
This has important consequences on the behavior of the system.
Assuming that a stationary state is reached, the probability density
$p_{\mathrm{st}}(q)$ to find
an atom at momentum $q$ is proportional to the density $\rho(q)$, and to
the lifetime $\tau(q)$:
\be
p_{\mathrm{st}}(q) = \frac{1}{\la \tau \ra} \rho(q) \tau(q),
\ee
with $\la \tau \ra = \int_0^{q_m} \dd q \rho(q) \tau(q)$.
For $\alpha < 1$, the distribution $p_{\mathrm{st}}(q)$ is no longer
normalizable since $\la \tau \ra$ is infinite, so that no stationary
distribution can be reached. This means that the dynamical distribution
of $q$ becomes more and more peaked around $q=0$ as time evolves,
allowing atoms to reach lower and lower momenta, or equivalently,
temperatures.

The same behavior can also be found in other physical contexts, like
the aging phenomenon in glassy systems. Actually, the kinetic mechanism at
play in the trap model of aging dynamics is precisely the same as for
laser cooling, although its physical origin is different.
Both models can actually be recast into the same formalism once expressed
in terms of lifetimes $\tau$ only.\cite{Bardou05}

\subsection{Convergence to the Gaussian distribution: 'finite-size' effects}

\subsubsection{Finite $N$ deviations from the Gaussian distribution}

In the previous section we presented mathematical results concerning the
sum of random variables. Obviously, these results relie on hypotheses that may
or may not be satisfied in physical systems.
Note also that the observation of Gaussian
distributions in physical systems does not mean that the CLT applies,
that is to say that the local variables are i.i.d.,
and that the second moment is finite:
Theorem \ref{CLT1} is only  an implication, not a equivalence.
Theorem \ref{CLT2} is an equivalence between the identity (\ref{condCLT})
and the Gaussian distribution, only assuming that the variables are i.i.d.

In most physical systems it is hard to know a priori whether these
hypotheses on the microscopic variables are valid or not.
Indeed, one most often studies fluctuations of macroscopic quantities
to gain a better understanding on the statistical properties at the
microscopic level. In this respect,
the observation of \textit{non}-Gaussian fluctuations is much more
interesting from the physicist's point of view: from the contra-positive of
the CLT, either the hypotheses are \textit{not} true
for the physical system considered, or the number of microscopic
random variables is not large enough to ensure the convergence towards
the limit distribution. Mathematics therefore makes observation of
non-Gaussian fluctuations much more interesting: it is generally the signature
of non-trivial physical phenomena in the system, leading either to correlations
or to broad distributions of the microscopic degrees of freedom.

However, before drawing such conclusions, one has to check that the number
of individual random variables is indeed large enough to use a limit theorem.
At first sight this condition could seem rather easy to fulfill,
but things might sometimes be more subtle than expected, as we shall
illustrate below. Let us imagine that we know that the
considered quantity is a sum of i.i.d.~random variables,
each with a finite variance. Mathematics and the CLT then tell us that the
\textit{limit} distribution of our quantity is a Gaussian distribution.
This is a strong result, with a clear-cut assumption: the variance is defined
or not. The problem is that the CLT does not in itself specify how fast the
convergence is, so that the number of (random) terms needed to actually
observe the convergence might be too large to be physically accessible.
Characterizing the convergence of the distribution of random sums towards
the normal distribution is thus an important issue. This is the aim of
the following theorem.\cite{Petrov75,Petrov95,Blinnikov}

\begin{theorem} \label{th-convergence}
Let $\{x_i \}_{i=1..N}$ be a set of $N$ random variables, independent and
identically distributed according to the distribution $p(x)$, with
$\langle x \rangle =m$, $\langle (x-m)^2 \rangle=\sigma^2$, $0<\sigma<\infty$,
and $\langle |x-m|^3 \rangle=\rho\,\sigma^3<\infty$. Then, for any integer
$N>0$, the cumulative distribution function $F_N(y)$ of the variable
\be
y = \frac{1}{\sigma\sqrt{N}} \left( \sum_{i=1}^N x_i - Nm \right)
\ee
satisfies the following inequality, called the Berry-Esseen inequality:
\be
\sup_y \left| F_N(y) - \frac{1}{\sqrt{2\pi}}\int_{-\infty}^y e^{-x^2/2}\, \dd x
\right| \le \frac{A\rho}{\sqrt{N}}
\ee
where $A$ is an absolute positive constant, independent of $N$ and
of the distribution $p(x)$.
\end{theorem}

Hence we see that the speed of convergence to the normal law is quite slow,
like $1/\sqrt{N}$.
In addition, the departure from the cumulative normal distribution
depends on the details of the distribution $p(x)$, through the ratio
$\rho$ of the third centered moment to the variance to the power $3/2$.
If $\rho$ is large, the convergence to the normal law can be so slow
that it may not be achieved on physically accessible scales, as illustrated
on the following example.

\subsubsection{Broad distribution effects in laws with finite but large
variance}

The lack of convergence to the normal law on experimentally accessible
scales has been observed by Da Costa and coworkers in experiments on
tunnel junctions.\cite{Dacosta00}
These authors studied the tunnel current measured through
metal-oxide junctions, in particular when the metal is Cobalt.
Their results show a strong heterogeneity of the local currents at the
film surface. The probability distribution of the local current
$i_\mathrm{loc}$ can be determined
experimentally and turns out to be non-Gaussian, and in good agreement
with a lognormal distribution, as predicted earlier by Bardou:\cite{Bardou97}
\be \label{lognormal}
p(i_\mathrm{loc}) = \frac{1}{\beta i_\mathrm{loc} \sqrt{2\pi}}
\exp \left[-\frac{1}{2\beta^2}
\left( \ln \frac{i_\mathrm{loc}}{i_0} \right)^2 \right],
\ee
where $\beta$ and $i_0$ are two junction-dependent parameters.
This lognormal distribution can rather easily be understood in terms
of Gaussian fluctuations of the height of the junction
surface.\cite{Bardou97,Dacosta00}

Now one can ask what is the distribution
of the total current through the junction.
Decomposing the area of the junction into small regions,
this current is the sum of local currents, assumed to be independent:
$I=\sum_{k=1}^N i_{\mathrm{loc},k}$. The total
current is therefore the sum of i.i.d.~random variables, and as the moments
of the lognormal distribution
are finite, the Central Limit Theorem applies: we thus expect
Gaussian fluctuations for the total current. Experimental results however
exhibit strong deviations from the Gaussian distribution: measured
distributions are indeed broad and skewed, even if the estimated value of $N$
is rather large.
The reason for this observation is found in the very slow convergence
toward the Gaussian distribution, which turns out to be out of reach in this
experimental system.\cite{Romeo03}
To check this point, we estimate the value of the parameter $\rho$ in
Theorem~\ref{th-convergence}, which requires to evaluate first the moments.
From Eq.~(\ref{lognormal}), the moments of the lognormal distribution are
given by
\be
\label{momlognorm}
\langle i_\mathrm{loc}^n \rangle = i_0^n e^{\beta^2 n^2/2}
\ee
Using the inequality $\langle |x-\langle x \rangle|^3 \rangle
\ge \langle (x-\langle x \rangle)^3 \rangle$, one finds
\be
\rho \ge \frac{\langle x^3\rangle-3\langle x\rangle\langle x^2\rangle
+2\langle x\rangle^3}{(\langle x^2\rangle-\langle x\rangle^2)^{3/2}}.
\ee
Together with Eq.~(\ref{momlognorm}), this last inequality leads to
\be
\rho \ge \frac{e^{3\beta^2}-3e^{\beta^2}+2}{(e^{\beta^2}-1)^{3/2}}
\ee
With the experimental value $\beta=1.84$,
one obtains the estimate $\rho \gtrsim 170$.
Assuming that the constant $A$ is of the
order of $1$ (it can be shown\cite{Petrov75} that $A \ge 1/\sqrt{2\pi}$),
one sees that Theorem~\ref{th-convergence} does not
impose any constraint on the convergence to the normal law until $N$
reaches values of the order of $\rho^2 \approx 3\times 10^4$.
Let us also note that the typical value of the current, 
that is the locus of the maximal probability, is 
\be
\label{ityp}
i_\mathrm{typ} = i_0 e^{-\beta^2},
\ee
yielding a ratio
\be
\frac{\langle i_\mathrm{loc} \rangle}{i_\mathrm{typ}} =
e^{\frac{3}{2}\beta^2} \approx 160.
\ee
So the average value is much larger than the typical value, which is
a standard property of broad distributions with finite first moment.
The typical value is experimentally
relevant as it corresponds to the most probable value measured in an
experiment.

This example shows that one should keep in mind that observing non-Gaussian
fluctuations might be the result of a lack of convergence toward the normal
distribution, due to effective broad distribution effects appearing
in finite-size systems.
The lognormal distribution precisely exhibits such properties if the
parameter $\beta$ is large enough (typically $\beta\gtrsim 1.5$).
Similar effects also appear in truncated power-law distributions, namely
distributions $p(x)$ such that $p(x)\propto x^{-1-\alpha}$ for
$X_\mathrm{min}<x< X_\mathrm{max}$ and with an exponential decay for
$x > X_\mathrm{max}$ (hence all moments are finite).
If $0<\alpha<2$ and $X_\mathrm{min} \ll X_\mathrm{max}$, 
one typically has $\rho \sim (X_\mathrm{max}/X_\mathrm{min})^{\alpha/2}$,
so that the convergence to the normal law is generically very slow.

\section{Sums and extreme values of dependent or non-identically
distributed variables}
\label{sect-non-iid}

\subsection{Statistics of random sums}

In this section we shall discuss the case of dependent or non
identically distributed random variables. These cases are very interesting
from the physicist viewpoint, and of course turn out to be quite subtle
on the mathematical side, apart from some simple cases.
We shall first explore what happens when the limits
of validity of Central Limit Theorem are reached.
While from a mathematical point of view, the hypotheses underlying the
CLT are either true or not, the boundaries are actually blurred
in physics. But after all, it is part of the physicist's \textit{savoir-faire}
to play with these blurred boundaries.

\subsubsection{Extension of the CLT to non-identically distributed variables}

Up to now we only considered the case of independent and identically
distributed random variables. Before including correlations, one can
consider the case when the variables are still independent, but no
longer identically distributed. In other words the joint probability of
$N$ random variables still factorizes, but the product involves different
marginal distributions:
\be \label{definid}
J_N(x_1,..., x_N)=\prod_{i=1}^N p_i(x_i).
\ee
The factorization property makes this case analytically tractable, and some
mathematical results are thus available.
It turns out that a slightly generalized form of the CLT can be applied,
provided that the so-called Lindeberg condition\cite{Feller} is fulfilled.
\begin{theorem}\label{lindeberg}
Let $\{ x_k \}_{k=1,\ldots,N}$ be a sequence of independent random variables 
with finite first and second moments and let $p_k(x_k)$ be the
probability distribution of $x_k$. In order that the probability distribution
of the normalized sum
\be
S_N=\frac{1}{b_N}\left(\sum_{k=1}^N x_k -a_N \right),
\ee
with
\be
a_N=\sum_{k=1}^N \langle x_k \rangle,\quad
b_N=\left( \sum_{k=1}^N (\langle x_k^2 \rangle - \langle x_k \rangle^2)
\right)^{\frac{1}{2}},
\ee
converges to the normal law, it is
necessary and sufficient that the Lindeberg condition holds for all
$\epsilon >0$, namely
\be
\label{eq-Lindeberg}
\lim_{N \rightarrow \infty} \frac{1}{b_N^2} \sum_{k=1}^N
\int_{|v|>\epsilon b_N} \dd v \, v^2 p_k\left(v+\langle x_k \rangle \right)=0.
\ee
\end{theorem}
An intuitive (although not fully correct) interpretation of the Lindeberg
condition is that the summands become infinitesimal with respect to the
sum in the large $N$ limit.
The Lindeberg condition is straightforwardly satisfied in the case of
i.i.d.~random variables. In this case, the condition reads
\be
\label{eq-Lindeberg2}
\lim_{N \rightarrow \infty} \frac{1}{b_1^2}
\int_{|v|>\epsilon b_N} \dd v \, v^2 p\left(v+\langle x \rangle \right)=0.
\ee
As $b_N$ goes to infinity when $N \rightarrow \infty$, condition
(\ref{eq-Lindeberg2}) is obviously satisfied.

Note that the Lindeberg condition implies that the standard deviation $b_N$
of the sum (or equivalently the variance $b_N^2$) diverges with $N$.
To show this, it is sufficient to show that the Lindeberg condition
cannot be fulfilled if $b_N$ is bounded.
In this case, since $b_N$ increases with $N$, it necessarily converges
to a finite limit. Hence the terms under the sum
in Eq.~(\ref{eq-Lindeberg}) asymptotically become independent of $N$.
As all terms are strictly positive (at least for small enough $\ve$),
the sum is larger than a positive bound,
and the limit in Eq.~(\ref{eq-Lindeberg}) cannot be zero,
since the prefactor $1/b_N^2$ also converges to a finite limit.

\subsubsection{$1/f$-noise: an example of non-identically distributed
variables}
\label{noise1}
If the Lindeberg condition is not satisfied,
one could observe other distributions
depending on the particular problem: there is no general or universal
behavior and one has to study each case separately.
A simple and explicit
example of such problems in physics was given by Antal and
coworkers.\cite{Antal01} These authors proposed a simple model for $1/f$-noise,
that we recast here in a slightly different form.
In this model, one considers a one-dimensional random signal $h_\ell$
parameterized by an integer $\ell=0,\ldots,N-1$.
This discretized signal may be interpreted either as a
time signal or as a spatial signal in a one-dimensional system.
Decomposing the signal $h_\ell$ using a discrete Fourier transform,
the complex Fourier amplitude $c_n$ associated with the wavenumber
$q_n=2\pi n/N$, $n=0,\dots,N-1$ reads:
\be
c_n = \frac{1}{\sqrt{N}} \sum_{\ell=0}^{N-1} h_\ell e^{-iq_n \ell}.
\ee

The basic assumption of this simple model
of $1/f$-noise is that the Fourier coefficients $c_n$, $n>0$,
are independent complex random variables with Gaussian distribution
and variance proportional to $1/n$:
\be \label{dist-cn}
p_n(c_n) = \frac{n\kappa}{\pi} e^{-n\kappa |c_n|^2}.
\ee
The value of $c_0$ is unimportant to the following, since we shall consider
fluctuations of $h_\ell$ around the empirical mean
$\overline{h}=N^{-1} \sum_{\ell=0}^{N-1} h_\ell = c_0/\sqrt{N}$.
Hence we define the 'roughness' of the signal as its empirical variance,
namely
\be \label{ERS}
E_N=\sum_{\ell=0}^{N-1} (h_\ell-\overline{h})^2,
\ee
This quantity, which fluctuates from one realization of the signal to
another, is a natural global observable in this context.
From Parseval's theorem, it may be conveniently reformulated as a function
of the Fourier amplitudes
\be \label{EFS}
E_N=\sum_{n=1}^{N-1} |c_n|^2,
\ee
so that we shall also call $E_N$ the integrated power spectrum
(or total energy) of the signal.

The integrated power spectrum $E_N$ can thus be written as a sum
of independent, but non-identically distributed variables $u_n \equiv |c_n|^2$.
From the Gaussian distribution of $c_n$ in the complex plane, it is
straightforward to show that the distribution of $u_n$ is given by
\be \label{dist-yn}
\tilde{p}_n(u_n) = n\kappa\, e^{-n\kappa u_n}
\ee
As a result, the variance of $u_n$ is equal to $\kappa^2/n^2$,
so that $\mathrm{Var}(E_N)=\sum_{n=1}^{N-1}\mathrm{Var}(u_n)$
converges to a finite limit as $N \rightarrow \infty$;
accordingly, Lindeberg's condition does not hold.

So as to deal with the infinite $N$ limit, it is convenient to introduce
a rescaled variable $\ve=(E_N-\la E_N\ra)/\sigma$,
where $\sigma^2$ is the infinite $N$ limit of the variance of $E_N$,
namely $\sigma^2 = \sum_{n=1}^{\infty} \kappa^2/n^2$.
Denoting as $\psi_N(\ve)$ the PDF of $\ve$,
we introduce the associated characteristic function $\chi_N(\omega)$
defined as
\be
\chi_N(\omega) = \int_{-\infty}^{\infty} \dd\ve\, \psi_N(\ve) e^{-i\omega \ve}.
\ee
Using the independence property of the variables $y_n$, one can express
$\chi_N(\omega)$ as the product of the characteristic functions
of $u_n$, $n=1,\ldots, N-1$. In the infinite $N$ limit,
$\chi_N(\omega)$ converges to the asymptotic characteristic function
$\chi_{\infty}(\omega)$ given by\cite{Antal01,BC06}
\be
\chi_{\infty}(\omega) = \prod_{n=1}^{\infty}
\left(1+\frac{i\omega}{n\sigma\kappa} \right)^{-1}
\exp\left(\frac{i\omega}{n\sigma\kappa}\right).
\ee
This expression can be transformed using the following relation,\cite{GR}
valid for any complex number $z \ne -1,-2,\ldots$,
\be
\Gamma(1+z) = e^{-\gamma z} \prod_{n=1}^{\infty} \frac{e^{z/n}}{1+\frac{z}{n}},
\ee
where $\Gamma(.)$ is the usual Euler Gamma function, and $\gamma=0.577\ldots$
is the Euler constant.
Computing the inverse Fourier transform of $\chi_{\infty}(\omega)$,
one finds that the limiting distribution $\psi_{\infty}(\ve)$ is precisely
a Gumbel distribution
\be
\psi_{\infty}(x)=g_1(x)=\exp[-(bx+\gamma)-e^{-(bx+\gamma)}],
\qquad b=\frac{\pi}{\sqrt{6}},
\ee
as introduced in Eq.~(\ref{eqGk}). This result
is rather striking since there is a priori here no relation with
extreme values statistics. Understanding whether the appearance of
the Gumbel distribution in the present context is a coincidence, or
if it unveils some deep connection between sums of random variables
and extreme values is the motivation of several
studies.\cite{Bramwell01b,Portelli02,Clusel04,Comtet07}
We shall discuss this point in details in Sect.~\ref{sect-sum-EVS}.

Finally, the example of the $1/f$-noise model illustrates how correlated
random variables (here the physical signal $h_{\ell}$) may in some simple
cases be converted into independent, but non-identically distributed
variables (the amplitudes $c_n$) through a Fourier transform.
Spatial correlations in the system then come from the
fact that the Fourier modes are extended objects.
The key point is that the global quantity of interest, namely the integrated
power spectrum $E$ can be expressed either as a function of the physical
signal $E=\sum_\ell (h_\ell -\overline{h})^2$ or as a function of the
Fourier amplitudes $E=\sum_n |c_n|^2$, thanks to Parseval's theorem.
Hence the statistics of the power spectrum can equivalently be considered
as a problem of sum of correlated variables or a problem of sum of
independent, but non-identically distributed random variables.
However, this is a rather specific class of problems, and in more general
cases, correlated variables cannot be converted into independent variables.
It is thus necessary to develop different approaches to tackle this issue.

\subsubsection{Correlated and identically distributed variables:
scaling arguments}

Generically, the case of correlated random variables
is very difficult from a mathematical point of view.
To the best of our knowledge,
there is no general criterion as Lindeberg's condition, to ensure the
applicability of the CLT for generic correlated variables.
However, some rigorous results exist in some specific cases.
For instance, it is known that the CLT still holds for a
particular case of correlation, the martingale
differences.\cite{Billingsley,Ibragimov,Feller2}
Convergence theorems for non-linear functionals of stationary Gaussian
sequences with power-law correlation have also been
obtained.\cite{Sun65,Taqqu77,Taqqu79,Dobrushin79,Rosenblatt81,Major81,Breuer83}
We shall come back to this issue in Sect.~\ref{sect-Taqq}.
In addition, propositions have been made recently concerning
extensions of the CLT to specific classes of
correlated variables using mathematical
concepts like deformed products,\cite{Baldovin07} or in relation
with Tsallis' non-extensive entropy.\cite{Tsallis06a,Tsallis06b,Tsallis06c}
However, this area is still a matter of
debate.\cite{Tsallis01,Pluchino06,Bouchet06,Schehr07}

From a less rigorous point of view, different strategies may be used,
from simple scaling arguments to more involved renormalization group
approaches (see Sect.~\ref{sect-RG}).
Let us start with a simple and intuitive physical argument.
Considering a large system of linear size $L$ in dimension $D$,
we assume that the microscopic degrees of freedom are typically correlated
over a length $\xi<L$.
Let us now imagine that we are interested in the statistics of a
particular global observable. The latter
can be expressed as a sum of local quantities computed on
subsystems with a linear size of the order of $\xi$.
Then as a first approximation, the quantities computed on two different
subsystems are statistically independent, so that the global observable
can be estimated as a sum of $N=(L/\xi)^D$ i.i.d.~random variables.
The main issue is now the behavior of the correlation length $\xi$
with the system size $L$. If $L/\xi \rightarrow \infty$ when
taking the thermodynamic limit $L \rightarrow \infty$, then the number $N$ of
independent terms in the sum goes to infinity.
At a heuristic level, one can apply the central limit 
theorem, leading to a Gaussian distribution for the sum.
In contrast, if the correlation length scales with the system size, that is
$\xi \sim L$, then $N$ remains finite and the central limit theorem does not
apply. In this case, the distribution obtained in the limit
$L \rightarrow \infty$ is generically not a Gaussian distribution.
Note that this heuristic argument is qualitatively consistent with the
Lindeberg condition: if the effective number of degrees of freedom
remains finite when $N \rightarrow \infty$, one expects that the variance
of the sum also converges, and Lindeberg's condition does not hold.

It is possible to make the above scaling argument sharper (though not fully
rigourous) using thermodynamic concepts as we shall now illustrate
in a ``magnetic language'' for definiteness --although nothing
is specific to magnetic systems here.
Consider a spin model with spins $\{s_i\}$, $i=1,\ldots,N$
interacting through a Hamiltonian $\mathcal{H}_0(\{s_i\})$.
In the presence of an external magnetic field $h$,
the Hamiltonian becomes
$\mathcal{H}(\{s_i\})=\mathcal{H}_0(\{s_i\})-h\sum_{i=1}^N s_i$.
The partition fonction $Z$ is given by:
\be
Z = \sum_{\{s_i\}} e^{-\beta \mathcal{H}(\{s_i\})}.
\ee
Then the successive derivatives of the free energy
$F(\beta,h)=-(1/\beta)\ln Z$ yield the cumulants of the total
magnetization $M=\sum_{i=1}^N s_i$:
\be
\la M^k \ra_c = \frac{\partial^k F}{\partial h^k}
= N \frac{\partial^k f}{\partial h^k},
\ee
where $f(\beta,h)=F(\beta,h)/N$ is the free energy per spin.
Since both $f(\beta,h)$ and $h$ are intensive quantities, it follows that
all the cumulants are proportional to $N$.
In particular, the average value $\la M \ra$ and the variance
$\sigma_N^2$ of $M$ scale with $N$. Let us thus write the variance
as $\sigma_N^2 = N \sigma^2$.
To check whether an asymptotic distribution exists, one needs to consider
the reduced variable $z$ defined as
\be
z = \frac{M-\la M \ra}{\sigma \sqrt{N}}.
\ee
The average value of $z$ vanishes by definition of $z$, and the cumulants
of order $k \ge 2$ are actually not affected by the shift by $\la M \ra$,
but only by the rescaling factor $1/\sigma\sqrt{N}$.
Accordingly, these cumulants are given by
\be
\la z^k \ra_c = \frac{N}{\sigma^k N^{k/2}} \,
\frac{\partial^k f}{\partial h^k}, \qquad k \ge 2.
\ee
Hence the cumulant of order $k \ge 2$ is proportional to $N^{1-k/2}$:
the second order cumulant remains finite when $N$ goes to infinity, while
higher order cumulants vanish in this limit.
This precisely means that the distribution of $z$
becomes Gaussian in the thermodynamic limit.

Note however that the above result implicitely assumes
that all partial derivatives of $f(\beta,h)$ with respect to $h$ are finite,
i.e., that $f(\beta,h)$ is regular.
Knowing whether the free energy per degree of freedom is well-defined
and regular in the thermodynamic limit is a difficult mathematical problem,
related in particular to the theory of large deviations.\cite{Ellis85}
From a physicist viewpoint, it is well-know that
at a second order critical point where correlations become strong,
the second order derivative of the free
energy with respect to $h$ (namely the susceptibility) diverges, and
the second order cumulant of $M$ no longer scales with $N$.
Thus the above thermodynamic argument breaks down, and the asymptotic
distribution can be distinct from a Gaussian, meaning that the effective
number of degrees of freedom remains finite.
This is due to the fact that the two-point correlation function
behaves as a power law, and that the only length scale in the problem
is the system size (apart from the microscopic length scale),
so that $\xi \sim L$.
Therefore, from a more general point of view, deviations from the Gaussian
distribution may be expected when the two-point correlation of a random
sequence behaves as a power law.

\subsubsection{Taqq's reduction theorem for Gaussian stationary sequences}
\label{sect-Taqq}

General theorems about correlated variables are seemingly difficult to obtain.
However, interesting limit theorems have been obtained for special classes
of correlated variables, namely non-linear functionals of stationary
Gaussian sequences.\cite{Sun65,Taqqu77,Taqqu79,Dobrushin79,Rosenblatt81,Major81,Breuer83}
A Gaussian sequence is characterized by a Gaussian joint probability
density\cite{Botet02}
(we consider here the case $\langle x_i \rangle=0$ for simplicity)
\be \label{Gauss-seq}
J_N(x_1,\ldots,x_N) = \frac{\sqrt{\mathrm{det} R}}{(2\pi)^{N/2}} 
\; \exp \left(-\frac{1}{2} \sum_{i,j=1}^{N} x_i R^{-1}_{ij} x_j \right),
\ee
where $R$ is a matrix of elements $R_{ij}$, and $R^{-1}_{ij}$ are the
matrix elements of the inverse matrix $R^{-1}$. If the matrix elements
are such that $R_{ij}=r(|i-j|)$ where $r(m)$ is a given function, then 
the Gaussian sequence is stationary.\footnote{The definition of a stationary
sequence in the general case is actually not obvious, and goes as follows.
Let $x_1,\ldots,x_k$ denote a set of $k$ random variables.
For any vector $\bm{i}(k)$ of $k$ integers, $\bm{i}(k)=(i_1,\cdots,i_k)$,
where $1\le i_1\le i_2 \cdots \le i_k \le k$, we note
\be
F_{\bm{i}(k)}(x_1,\ldots,x_k)=\mathrm{Prob}(x_{i_s}<x_s, 1\le s \le k).
\ee 
The sequence $\{x_k \}$ is stationary if the probability
$F_{\bm{i}(k)}(x_1,\ldots,x_k)$ stays invariant if the vector $\bm{i}(k)$
is translated by any vector $\bm{m}=(m,\ldots,m)$, with $m$ any positive
integer:
\be
F_{\bm{i}(k)+\bm{m}}(x_1,\ldots,x_k)=F_{\bm{i}(k)}(x_1,\ldots,x_k).
\ee
Note in particular that i.i.d.~random variables form a stationary sequence.}
In this case, the marginal probability for each random variable is a normal
distribution of variance $r(0)$, and the two-point correlation
function reads $\langle x_i x_{i+m} \rangle =r(m)$.

In the following, we focus on Gaussian stationary sequences
$x_1,\ldots,x_N$ such that the correlation function $r(m)$
decays at large distance as a power law, $r(m) \sim m^{-\alpha}$
(note that for a rigorous definition of this power law behavior, the limit
$N \rightarrow \infty$ should be taken before the limit
$m \rightarrow \infty$).
We then introduce a sequence of variables $y_1,\ldots,y_N$ through
\be
y_i = \psi(x_i), \qquad i=1,\ldots,N,
\ee
where $\psi(x)$ is a regular function taking real values, and such that
\bea \label{cdt-psi1}
\int_{-\infty}^{\infty} \dd x \, \psi(x) \, e^{-x^2/2} =0 \\
\int_{-\infty}^{\infty} \dd x \, \psi(x)^2 \, e^{-x^2/2} < \infty
\label{cdt-psi2}
\eea
Then $\psi(x)$ can be expanded over the basis of Hermite polynomials
$H_j^*(x)$, namely
\be
\psi(x) = \sum_{j=1}^{\infty} c_j H_j^*(x).
\ee
Hermite polynomials are defined as $H_1^*(x)=x$, $H_2^*(x)=x^2-1$,
and the recursive relation $H_{j+1}^*(x)=xH_{j}^*(x)-jH_{j-1}^*(x)$ for $j\ge 2$.
Then $\psi(x)$ is said to have Hermite rank $m^*$ if the first non-vanishing
coefficient in the expansion is $c_{m^*}$, that is $c_j=0$ for
$j<m^*$ and $c_{m^*} \ne 0$.\cite{Taqqu75,Breuer83}

The following theorem\cite{Taqqu79,Breuer83,Botet02}
then holds for the sum of the variables $y_i$.

\begin{theorem} \label{th-Taqqu}
Let $\{x_i\}_{i=1,\ldots,N}$ be a Gaussian stationary sequence with
correlation function decaying at large distance as $r(m) \sim m^{-\alpha}$.
Define the sequence $y_i = \psi(x_i)$ with $\psi(x)$ a real function with
Hermite rank $m^*$. If $\alpha<1/m^*$, the asymptotic distribution of the sum
$S_N=\sum_{i=1}^N y_i$ exists, and is non-Gaussian if $m^*>1$, while it is
Gaussian for $m^*=1$.
In the opposite case $\alpha>1/m^*$, the Gaussian distribution
is recovered.
\end{theorem}

The first part of the theorem is known as Taqqu's reduction
theorem.\cite{Botet02} The corresponding non-Gaussian limit
distributions are known through their cumulants,
given by multiple integrals.\cite{Dobrushin79}
Note that an explicit example with $m^*=2$ and $\alpha<1/m^*$
was given by Rosenblatt\cite{Rosenblatt61} before Theorem~\ref{th-Taqqu}
was proven.
Moreover, this theorem shows that, at least for stationary
Gaussian sequences, rather strong correlations can be included without
affecting the limit distribution, which remains Gaussian.

From a more general perspective, it seems that
when considering generic classes of strongly correlated variables,
almost any ``reasonable'' function could be a particular limit function,
so that trying to classify them is hardly possible.
For instance, a continuous family of limit functions is obtained
in the simple $1/f^{\alpha}$-noise model\cite{Racz02} of correlated random
signals.
Still, from a physicist point of view,
the knowledge gained from the renormalization group approach in
statistical mechanics might suggest that physically relevant classes
of strongly correlated random variables may be
organized in kinds of universality classes (a notion somehow close to
that of basin of attraction appearing in convergence theorems).
In this rather optimistic picture, one could a priori guess what kind
of probability distribution is related to a particular physical problem,
based on general symmetry and dimensionality properties.
However, even assuming that such a ``gallery'' of asymptotic
distributions could be defined, it is presently far
from being quantitatively completed. Indeed, most of the known results
in the physics literature are obtained through perturbative expansions,
\cite{Binder87,Diehl87} and the asymptotic distributions are not known
exactly in most cases. Moreover, the above gallery of distributions
is continuously expanding, through the development of non-equilibrium
statistical physics.\cite{Racz03}

\subsubsection{The renormalization group approach}
\label{sect-RG}

The renormalization group procedure has had an enormous impact in physics,
specifically in the study of critical phenomena,
but also in diverse fields of physics like field theory, or the study
of disordered systems.\cite{Wiese03}
The main idea of the renormalization group approach is to coarse-grain
step by step the description of the system, while conserving the
thermodynamic properties. In more mathematical terms, this means that
the original random variables describing the microscopic degrees of freedom
are coarse-grained iteratively into ``mesoscopic'' effective random variables,
and that the statistical properties of the sum (for instance the total
magnetization in a magnetic system) is preserved. This is due to the fact that
the transformation conserves the partition function; since the logarithm
of the partition function generates the magnetization cumulants,
it follows that this distribution is conserved
through the renormalization procedure.

In the following, we shall briefly illustrate on a standard solvable example,
the decimation of the Ising chain,\cite{Fisher75} how a
renormalization group approach may be used to determine the
asymptotic distribution of a sum of correlated variables
(in the same spirit, an exact renormalization procedure
can be performed in the one-dimensional XY-model.\cite{Droz78})
Note that in practice, examples that can be solved
exactly through a renormalization group calculation can most often
be solved by other more direct methods. Yet, various approximation schemes
exist in more complicated situations, and provide valuable
insights into the behavior of the system.

Let us also mention that we are not mostly interested
here in the critical point
of the model, which is a zero-temperature critical point, but rather
by the statistics of the magnetization away from the critical point,
that is at finite temperature. In this case, the intuitive scaling
argument presented above suggests that since the correlation length
$\xi$ is finite, the effective number $N/\xi$ of degrees of freedom
($D=1$ here) diverges in the thermodynamic
limit, so that one expects to recover a Gaussian distribution.
The decimation procedure allows us to obtain this result explicitely,
as through the renormalization process, the joint distribution converges
to a factorized distribution (and moments are finite).

The partition function of the Ising chain (or one-dimensional Ising model)
with an even number $N$ of sites, and periodic boundary conditions,
is given by
\be
\mathcal{Z} = \sum_{\{s_i\}} \exp\left(-\beta\sum_{i=1}^N 
\mathcal{H}_{i,i+1}(s_i,s_{i+1})\right),
\ee
where the local Hamiltonian $\mathcal{H}_{i,i+1}(s_i,s_{i+1})$ associated
with the link $(i,i+1)$ is given by
\be
\mathcal{H}_{i,i+1}(s_i,s_{i+1}) = -K s_i s_{i+1} - \frac{h}{2}(s_i+s_{i+1})
+c.
\ee
Note that the role of $s_i$ and $s_{i+1}$ have been symmetrized for later
convenience, and that a constant term $c$ has been added. This constant
term is irrelevant at this stage and could be set to zero,
but such a term will be generated by the renormalization procedure,
and it is useful to include it from the beginning.

The basic idea of the decimation procedure is to perform, in the partition
function, a partial sum over the spins of --say-- odd indices in order to
define renormalized coupling constants $K'$ and $h'$.
Then summing over the values of the spins with even indices yields the
partition function $\mathcal{Z}'$ of the renormalized model,
which is by definition of the renormalization procedure equal to the
initial partition function $\mathcal{Z}$.
To be more explicit, one can write $\mathcal{Z}$ as
\be
\mathcal{Z} = \sum_{\{s_{2j}\}} \sum_{\{s_{2j+1}\}} 
\exp\left(-\beta\sum_{i=1}^N \mathcal{H}_{i,i+1}(s_i,s_{i+1})\right)
\ee
and rewrite the above equality in the following form:
\be
\mathcal{Z} = \sum_{\{s_{2j}\}}
\exp\left(-\beta\sum_{j=1}^{N/2} \mathcal{H}_{j,j+1}'(s_{2j},s_{2j+2})\right)
\ee
where $\mathcal{H}'(\{s_{2j}\})=\sum_{j=1}^{N/2}\mathcal{H}_{j,j+1}'(s_{2j},s_{2j+2})$
is the renormalized Hamiltonian, defined by
\be
\exp\left[-\beta \mathcal{H}'(\{s_{2j}\})\right] = 
\sum_{\{s_{2j+1}\}} \exp\left( -\beta \sum_{i=1}^N \mathcal{H}_{i,i+1}(s_i,s_{i+1})\right).
\ee
This last relation is satisfied if, for any given $j=1,\ldots, N/2$,
and any given values of $s_{2j}$ and $s_{2j+2}$,
\bea \nonumber
&&\exp\left(-\beta \mathcal{H}_{j,j+1}'(s_{2j},s_{2j+2})\right)\\
&&\quad = \sum_{s_{2j+1}=\pm 1}
\exp\left(-\beta \left[\mathcal{H}_{2j,2j+1}(s_{2j},s_{2j+1})+
\mathcal{H}_{2j+1,2j+2}(s_{2j+1},s_{2j+2})\right]\right).
\eea
Assuming that $\mathcal{H}_{j,j+1}'(s_{2j},s_{2j+2})$ takes the form
\be
\mathcal{H}_{j,j+1}'(s_{2j},s_{2j+2}) = -K's_{2j} s_{2j+2}
-\frac{h'}{2} (s_{2j}+s_{2j+2})+c',
\ee
one obtains, with the notation $s=s_{2j+1}$,
\bea \label{eq-renorm1}
&&\exp\left(\beta K' s_{2j} s_{2j+2}+\frac{\beta h'}{2}(s_{2j}+s_{2j+2})-\beta c'\right)\\ \nonumber
&=& \sum_{s=\pm 1} \exp\left(\beta K s_{2j} s+\beta K s s_{2j+2} 
+ \frac{\beta h}{2}(s_{2j}+s_{2j+2}) + \beta h s -2\beta c\right).
\eea
Introducing the reduced variables
\be
u=e^{-4\beta K}, \qquad v=e^{-2\beta h}, \qquad \omega = e^{-4\beta c},
\ee
Eq.~(\ref{eq-renorm1}) yields the following coupled recursion relations:
\be
u' = \frac{u(1+v)^2}{(u+v)(1+uv)}, \quad
v' = \frac{v(u+v)}{1+uv}, \quad
\omega' = \frac{\omega^2}{uv^2} (1+v)^2 (u+v)(1+uv).
\ee
One sees that the evolution of $u$ and $v$, which correspond to the physical
coupling constants, is actually decoupled from the evolution of $\omega$ which
encodes the (apparently useless) constant term in the
Hamiltonian.\footnote{Note however that the evolution of $\omega$ under
renormalization still contains some useful information, as one can compute
from it the free energy at any temperature.}
When starting from a finite temperature so that $u>0$ initially,
iterations of the renormalization recursion relations leads
to one of the fixed point $u_0=1$, $v_0 \ge 0$
corresponding to a system with coupling
constant $K=0$ and arbitrary external field $h$.
Since $K=0$, the spins are independent random variables, so that
the CLT applies (the second moment is finite). As the distribution
is the same as that of the original system which included a
coupling between the spins, one concludes that the magnetization
in the correlated system also has a Gaussian distribution
in the thermodynamic limit.
Note however that one has to assume that the order of the two
limits $N \rightarrow \infty$ and infinite number of iterations of
the renormalization group can be exchanged, which is not necessarily obvious.

Finally, let us note that the fixed points we have considered here,
where the spins become independent random variables, is called
a ``trivial'' fixed point. Generally speaking, what is more interesting
from the physicist's viewpoint is the so-called ``critical'' fixed point,
in which the spins become highly correlated. 
In the Ising chain however, it corresponds to $u=0$ and $v=1$,
that is to infinite coupling $K$, or zero temperature, and zero field $h$.
The absence of finite temperature phase transition is
generic in one-dimensional equilibrium systems with short-range
interactions.\cite{Lieb66}

\subsection{Statistics of extreme values}

\subsubsection{Extreme values of non-identically distributed independent
variables}

As in the case of statistics of sums of random variables, losing the
i.i.d.~property leads to much weaker results concerning the limit
distributions of extremes.
In this section, we focus on sequences of 
independent random variables $\{ x_i\}$, with different marginal
cumulative probability distributions $F_i$.  This case has deserved
attention recently due to its practical interest, for example in the
study of extreme climatic events, where climate changes lead to a
modification of the underlying statistics,\cite{Naveau05,Naveau06}
or in application of ideas of extreme value statistics in evolving
risk insurance.\cite{Potters}
The distribution of the maximum then satisfies 
\be \label{maxnonid}
F_N^{\mathrm{max}}(z) \equiv \mathrm{Prob}(\max(x_1,\ldots,x_N) \le z) =
\prod_{i=1}^N F_i(z),
\ee
which is an extension of Eq.~(\ref{eqext1}).
Note that in practical applications however, the marginal distributions
are generally unknown, which makes the above equation essentially useless
for practical purposes.

If the random variables are strongly non-identical,
the asymptotic distribution of the maximum could be 
any probability distribution, as shown on a simple example below.
Hence, the identification of asymptotic laws for extremes of
non-identical random variables has to be done case by
case,\cite{Falk,Galambos,Coles,Gyorgyi03}
and is beyond the scope of the present article.

Some simple understanding of these issues can be gained using
an interesting example inspired by Falk and coworkers.\cite{Falk}
It makes use of the following simple property:
if $F_0$ is a cumulative distribution function,
then for any real $\gamma>0$, the function $F_0^\gamma$ is also a cumulative
distribution function.
Hence a simple way to generate non-identical random variables is to choose
a cumulative distribution function $F_0$ and a set of $N$ number
$\gamma_i>0$, and to define a sequence of independent
random variables $\{x_i\}$ with cumulative distribution
$F_i(z)=F_0(z)^{\gamma_i}$, $i=1,\ldots,N$.
Relation~(\ref{maxnonid}) then leads to
\be
\mathrm{Prob}(\max(x_1,\ldots,x_N) \le z)
= F_0(z)^{\sum_{i=1}^N \gamma_i}.
\ee 
It is then clear that the asymptotic law is controled by the behavior
of the sum $S_N = \sum_{i=1}^N \gamma_i$ when $N \rightarrow \infty$.
If $S_N$ converges to a finite limit $S_{\infty}$, the asymptotic
cumulative distribution is simply given by $F_0(z)^{S_{\infty}}$,
which can be any distribution since $F_0$ is arbitrary.

In contrast, if $S_N \rightarrow \infty$ when $N \rightarrow \infty$,
it is rather easy to show that the standard extreme value distributions
(namely, the Gumbel, Fr\'echet and Weibull ones) are obtained.
The asymptotic distribution is selected among the three possible ones
according to the function $F_0$, with the same criteria as those
presented in Sect.~\ref{iidext1}. For the sake of simplicity,
we illustrate this result on the example of the Gumbel
distributions, but the same argument holds for Fr\'echet and Weibull
distributions.
Defining the function $\zeta(z)$ through $F_0(z)=1-\exp(-\zeta(z))$,
one then has from Eq.~(\ref{maxnonid})
\be
F_N^{\mathrm{max}}(z) = \left( 1-e^{-\zeta(z)}\right)^{S_N}.
\ee
Since we focus on the Gumbel case, we consistently assume that
$\zeta(z) \approx \lambda z^{\nu}$ when $z\rightarrow \infty$,
with $\lambda>0$ and $\nu>0$
(although special cases of bounded variables could also be considered
within the Gumbel class, as mentioned in Sect.~\ref{iidext1}).
One would then like to know whether there exists a sequence of reals numbers
$a_N$ and $b_N$ such that $F_N^{\mathrm{max}}(a_N+b_N x)$ converges
to the Gumbel cumulative distribution $H_g(x)=\exp(-e^{-x})$.
In the large $N$ limit, we write $\zeta(a_N+b_N x)$ as
\be
\zeta(a_N+b_N x) \approx  \lambda (a_N+b_N x)^{\nu}
= \lambda a_N^{\nu} \left( 1+\frac{b_N}{a_N}x \right)^{\nu}.
\ee
Let us assume that $b_N/a_N \rightarrow 0$ when $N \rightarrow \infty$,
and self-consistently verify this assumption afterwards.
Expanding to first order in $b_N/a_N$, we have
\be
\zeta(a_N+b_N x) = \lambda a_N^{\nu} + \lambda\nu b_N a_N^{\nu-1}x
+\mathcal{O}\left( b_N^2 a_N^{\nu-2}\right).
\ee
We choose $a_N$ and $b_N$ such that
\be
\lambda a_N^{\nu} = \ln S_N, \qquad  b_N a_N^{\nu-1} = \frac{1}{\lambda\nu}.
\ee
Then, $a_N = (\lambda^{-1} \ln S_N)^{1/\nu}$ diverges with $N$, and
\be
b_N^2 a_N^{\nu-2} = \frac{1}{(\lambda\nu)^2\, a_N^{\nu}} 
\sim \frac{1}{\ln S_N} \rightarrow 0
\ee
when $N \rightarrow \infty$. One then has
\be
\zeta(a_N+b_N x) = \ln S_N + x +\mathcal{O}\left(\frac{1}{\ln S_N}\right).
\ee
One also verifies that the choice of $a_N$ and $b_N$ is consistent with
the assumption that $b_N/a_N \rightarrow 0$ when $N \rightarrow \infty$,
since $b_N/a_N=1/(\lambda\nu a_N^{\nu})$.
Plugging these results into the cumulative distribution
$F_N^{\mathrm{max}}$, one obtains
\be
F_N^{\mathrm{max}}(a_N+b_N x) =
\left[ 1-\frac{1}{S_N} \exp\left( -x+
\mathcal{O}\left(\frac{1}{\ln S_N}\right)\right)\right]^{S_N}.
\ee
Taking the limit $N\rightarrow\infty$, one finds, as $S_N\rightarrow\infty$,
\be
F_N^{\mathrm{max}}(a_N+b_N x) \rightarrow \exp\left(-e^{-x}\right)
\ee
which is nothing but the cumulative Gumbel distribution.
Hence one recovers for this specific class of non-identically distributed
and independent random variables the standard behavior of i.i.d.~variables
belonging to the Gumbel class. Note however that the convergence to
the asymptotic distribution may be quite slow, due to the logarithmic
dependence on $S_N$.

\subsubsection{Extreme of correlated variables}

The case of extreme value statistics of correlated random variables
has deserved attention in recent years, due to its application to
physical situations such as the Random Energy Model
(REM),\cite{Derrida81,Derrida81b,Carpentier} fluctuating
interfaces,\cite{Gyorgyi03,Comtet04,Comtet05,Lee05,Schehr06,Gyorgyi07}
directed polymers,\cite{Johansson00,Krapivsky00,Dean01}
Burgers turbulence,\cite{BouchMez97,Noullez02},
freely expending gases in one-dimension,\cite{Bena07}
biological evolution of quasispecies\cite{Krug03}
or applied statistics and climatology.\cite{Schlather03,Cooley06}
A slightly different issue has also been addressed recently,
namely the maximum value of a time signal with respect to the initial
value.\cite{Burkhardt07}
Rigourous mathematical treatments are only available
in a very few cases (and more specifically the Gaussian
cases\cite{Galambos,BaldassarriTh,Falk,Gyorgyi03}),
but approximate treatments, using physicist's tools such as replica
trick\cite{BouchMez97} or functional renormalization group
approaches,\cite{Carpentier} give some indications on the consequence
of correlations.

Let us first show using an example\cite{Falk} that, as for
non-identically distributed variables,
any probability distribution could be seen as an asymptotic law for extreme 
of correlated variables. To that purpose consider a set of i.i.d.~random
variables $\{y_i\}$ described by an arbitrary distribution, and
a random variable $v$, independent of the variables $y_i$,
described by a cumulative distribution $F(v)$. Now define a new set of random
variables $x_i$ through $x_i=v+y_i$ for all $i$. If there exists a sequence 
of real numbers $\{ a_N \}$ such that
\be \label{asex}
\forall \epsilon>0,\, \lim_{N\rightarrow \infty}
\mathrm{Prob}(|\max(y_1,\ldots,y_N) -a_ N| >\epsilon) =0, 
\ee
then in the limit $N \rightarrow \infty$, for any real $z$: 
\bea \nn
\mathrm{Prob}(\max(x_1,\ldots,x_N)-a_N \le z)
&=& \mathrm{Prob}(v+\max(y_1,\ldots,y_N)-a_N \le z)\\
&\rightarrow& \mathrm{Prob}(v\le z) = F(z).
\eea
due to Eq.~(\ref{asex}). This is the case for instance if $y_i$ is uniformly
distributed over a given interval.
We then have an extreme of correlated random variables which is
distributed according to an arbitrary probability distribution. 
This means that as in the case of non-identically distributed random
variables, the notion of classes, or of basin of attraction, is less
useful that in the i.i.d.~case.

Rigorous results can be derived in the particular case of stationary
Gaussian sequences of correlated random variables,\cite{Galambos,BaldassarriTh}
as defined in Eq.~(\ref{Gauss-seq}), for which $\langle x_i \rangle=0$
for all $i$, and the two-point correlation function depends only on $|i-j|$,
namely $\langle x_i x_{i+m} \rangle =r(m)$.
Note that the basin of attraction of these theorems is therefore quite
small compared to the ones of the CLT or even the generalized CLT.
This is a direct consequence of the complexity induced by the loss
of the independence between random variables.

The first kind of results deals with the case
of constant correlation,
such that  $\langle x_i x_{i+m} \rangle =r_N $ for all $m$ and $i$
(the value $r_N$ depending on the sample size $N$ only).
The extreme distribution depends in this case
on the behavior of $r_N$ as a function of $N$.\cite{Galambos}

\begin{theorem} \label{thEVScorrel1}
Let $y_N=\max(x_1,\ldots,x_N)$ be the maximum of $N$ elements of a Gaussian
sequence $\{x_k\}_{k=1,\ldots,N}$ with zero mean, unit variance,
and constant correlation $r=r_N$. Let 
\be \label{aNbN-EVS1}
a_N=\frac{1}{b_N}-\frac{1}{2} b_N \left( \ln \ln N + \ln 4\pi \right),
\qquad b_N=(2 \ln N)^{-1/2}.
\ee
If, as $N \rightarrow \infty$, $r_N \ln N$ converges to a finite value $\tau$,
then the rescaled variable $z=(y_N-a_N)/b_N$ has a limit distribution.
\begin{itemize}
\item If $\tau =0$, the limit distribution of $z$ is the
Gumbel distribution as in the i.i.d.~case.
\item If $\tau>0$, the limit distribution of $z$ is
the convolution of a translated Gumbel distribution
$\exp\left(-(z+\tau)-e^{-(z+\tau)} \right)$, and a Gaussian distribution with
zero mean and variance $2\tau$.
\end{itemize}
If $\lim_{N \rightarrow \infty} r_N \ln N=\infty$, then
the variable $z'=(y_N-a_N\sqrt{1-r_N})/\sqrt{r_N}$ has a limit distribution,
which is the normal distribution with zero mean and unit variance.
\end{theorem}

We see that only in the case of weak enough correlation (first case),
the asymptotic distribution remains the same as in the i.i.d.~case
(because the Gaussian distribution of the variables $x_i$ is in the basin
of attraction of the Gumbel asymptote).
Interestingly, another connection between extreme values and sums
seems to appear here, since one finds a Gaussian distribution
(typical of sums of i.i.d. random variables) as asymptotic
distribution of extreme values of correlated random variables.
We shall come back to this issue in Sect.~\ref{sect-sum-EVS}.

The second kind of results applies to the perhaps more realistic case
(at least from a physicist point of view) of a correlation $r(m)$
decreasing with the distance $m$. The limit distribution of the
maximum of the sequence depends on how fast the correlation decays, the
transition between the two regimes being once again a logarithmic
decay.\cite{Galambos,Carpentier}

\begin{theorem} \label{thEVScorrel2}
Let $y_N=\max(x_1,\ldots,x_N)$ be the maximum of $N$ elements of a Gaussian
sequence $\{x_k\}_{k=1,\dots,N}$ with zero mean, unit variance,
and correlation $r(m)=\langle x_k x_{k+m} \rangle$, independent of $k$.
If, as $m \rightarrow \infty$, $r(m) \ln m$ converges to a finite value
$\tau$, then the rescaled variable $z=(y_N-a_N)/b_N$ has a limit
distribution when $N \rightarrow \infty$, where $a_N$ and $b_N$ are given
in Eq.~(\ref{aNbN-EVS1}).
\begin{itemize}
\item If $\tau=0$, the correlation
does not affect the limit distribution of the maximum, which is the Gumbel
distribution as in the i.i.d.~case.
\item If $\tau>0$, the asymptotic distribution of the maximum is
the convolution of a translated Gumbel distribution
$\exp\left(-(z+\tau)-e^{-(z+\tau)} \right)$, and a normal distribution
with zero mean and variance $2\tau$.
\end{itemize}
If $\lim_{m \rightarrow \infty} r(m) \ln m=\infty$ and
$\lim_{m \rightarrow \infty} r(m) (\ln m)^{1/3}=0$,
then the variable $z'=(y_N-a_N\sqrt{1-r(N)})/\sqrt{r(N)}$
has a limit distribution, which is the normal distribution
with zero mean and unit variance.
\end{theorem}

These results were also derived using functional RG calculations, and
applied to the study of the glass transition in the REM with correlated
Gaussian potential.\cite{Carpentier}

\subsection{Relation between statistics of sums and of extreme values}
\label{sect-sum-EVS}

\subsubsection{Are there extreme values hidden in 1/f-noise?}

Let us come back to the $1/f$-noise problem studied in the previous
section, where it was found that the PDF of the integrated
power spectrum (or total energy) given by Eqs.~(\ref{ERS}), (\ref{EFS}),
is the Gumbel distribution $G_1$, while there is a priori no
connection with extreme values in this problem.
A rather natural idea would be that extreme values may be somehow ``hidden''
in the computation of certain types of random sums, in the sense that
a few terms may dominate the sum. As discussed before, this is the case
when considering broad distributions of identically distributed variables
--although the distribution of the sum is not in this case an extreme value
distribution. In the present situation, the distributions of the individual
variables are not broad --the distribution of $u_n=|c_n|^2$ is exponential,
as seen in Eq.~(\ref{dist-yn}).
Still, one might imagine that due to the dispersion of the
variances of the different variables $|c_n|^2$, some of the variables
(the ones with small $n$'s) may dominate the sum, and lead to
distributions similar to that found in the context of extreme statistics.
However, attempts to identify such dominant contributions in problems
essentially similar to the $1/f$-noise, concluded that all terms
in the sums are necessary to account for the observed probability
distributions. To be more specific, such a study was done in the context of
the XY-model, where contributions of the longest wavelength modes were
studied,\cite{Portelli02,PortelliTh} and in the Ising model, where the
contribution of the largest connected cluster of --say-- up spins was
identified.\cite{Clusel04} In light of these results, it seems clear that
the relation with extreme value statistics, if it exists, is more subtle, and
does not come from the dominant contribution of a few terms.

\subsubsection{Mapping an extreme value on a sum}

Actually, the situation becomes clearer when one takes another perspective.
Up to now, we have been looking for extreme values hidden in problems of sums.
Let us now take a different point of view: could we instead look for
sums hidden in extreme values? Although this question might seem a bit
strange at first sight, its answer actually contains the explanation
of the puzzling relation between statistics of extreme values and
statistics of some classes of random sums,
analogous to the $1/f$-noise problem.

To see how sums may emerge out of extreme values,\cite{Bertin05,BC06}
let us consider a set of $N$ i.i.d.~random variables $x_n>0$ ($1\le n\le N$),
drawn from a distribution $p(x)$.
Let us order these variables by relabelling them into $y_n$
in such a way that $y_1\ge y_2\ge\dots\ge y_N$. Formally, this means that
there exists a permutation $\sigma$ over the integers $1,\ldots,N$ such that
$y_n=x_{\sigma(n)}$. By definition, $y_1=\max(x_1,\ldots,x_N)$,
and similarly $y_k$ is the $k^{\rm th}$ largest value
among the set $\{x_n\}$.
Once properly rescaled, the variable $y_1$ necessarily follows one of the
three classes of extreme value statistics given in Theorem~\ref{th-EVS1}.

Then, one can introduce the differences between the ordered variables,
defining $v_n$ as
\be \label{mapping}
v_n = y_n - y_{n+1} \qquad (1\le n\le N-1), \qquad v_N=y_N.
\ee
It results that the largest value $y_1$ among the $x_n$'s
can be rewritten as
\be
\max(x_1,\ldots,x_N) \equiv y_1 = \sum_{n=1}^N v_n.
\ee
Similarly, the  $k^{\rm th}$ largest value $y_k$ can be expressed as
\be
y_k = \sum_{n=k}^N v_n.
\ee
Accordingly, it turns out that extreme values can be quite naturally
expressed as sums of random variables.
Note that up to now, we have only introduced a formal procedure to recast
an extreme value into a sum. To be more quantitative, we need to determine
the statistical properties of the variables $v_n$.
To this aim, let us introduce the joint distribution
$\Phi_{k,N}(v_k,\dots,v_N)$ of the variables $v_k,\dots,v_N$.
This joint distribution is formally defined as:
\begin{eqnarray}
\Phi_{k,N}(v_k,...,v_N)&=&N!\int_0^{\infty} \dd y_N p(y_N)
\int_{y_N}^{\infty} \dd y_{N-1} p(y_{N-1}) ...\int_{y_2}^{\infty}
\dd y_1 p(y_1) \nn \\
&& \qquad \qquad \times
\delta(v_N-y_N) \prod_{n=k}^{N-1}\delta(v_n-y_n+y_{n+1}).
\end{eqnarray}
From this definition, it is straightforward to show that\cite{BC06}
\be \label{PhikN}
\Phi_{k,N}(v_k,...,v_N)=\frac{N!}{(k-1)!}\; \tilde{F}\left(\sum_{i=k}^N
v_i\right)^{k-1} \prod_{n=k}^{N}p\left(\sum_{i=n}^N v_i\right),
\ee
where the function $\tilde{F}(z)$ is given by
\be
\tilde{F}(z)\equiv \int_z^{\infty}\dd y \ p(y).
\ee
It is now convenient to perform a shift of indices by introducing
$u_n=v_{n+k-1}$, leading to redefine the distribution
$\Phi_{k,N}(v_k,...,v_N)$ into $\tilde{\Phi}_{k,N'}(u_1,...,u_{N'})$, with
$N' \equiv N-k+1$. The expression of $\tilde{\Phi}_{k,N'}(u_1,...,u_{N'})$
is immediately deduced from Eq.~(\ref{PhikN}):
\be \label{PsikN}
\tilde{\Phi}_{k,N'}(u_1,...,u_{N'})
=\frac{(N'+k-1)!}{(k-1)!}\; \tilde{F}\left(\sum_{i=1}^{N'}
u_i\right)^{k-1} \prod_{n=1}^{N'} p\left(\sum_{i=n}^{N'} u_i\right),
\ee
From the very definition of the variables $u_n$, it is clear that
the sum $\sum_{n=1}^{N'} u_n$, once properly rescaled, converges
to one of the asymptotic extreme value distributions $g_k$, $f_{k,\mu}$
or $w_{k,\mu}$ defined in Eqs.~(\ref{eqGk}), (\ref{eqFk}) and (\ref{eqWk}).

It is interesting to note that when describing experimental
or numerical data, generalized extreme value distributions, taking $k$
as a real and positive fitting parameter, are often
considered.\cite{Bramwell01b,Brey05}
The question then naturally arises to know whether one could give
a precise meaning (other than phenomenological fitting functions) to such
generalized extreme value distributions.

\subsubsection{Generalized extreme value distributions for sums
of non-i.i.d.~variables} \label{generalization}

It turns out that the answer to this question is rather
straightforward, given the results presented in the previous section.
Indeed, from Eq.~(\ref{PsikN}), one sees that the integer $k$ is now
a simple parameter of the distribution, and that the distribution
$\tilde{\Phi}_{k,N'}(u_1,...,u_{N'})$ could easily be generalized to
non-integer values of $k$, provided that factorials are replaced
by Gamma functions.
In what follows, we shall denote $k$ as $a$ whenever it is not restricted
to integer values.
Note that one now needs to consider $N'$ as an integer rather than $N$.
In other words, the direct mapping back to the problem of extreme values
of i.i.d.~random variables is no longer possible for non-integer $k$
(see however the discussion in Sect.~\ref{sect-back}).
Accordingly, it is necessary to perform an independent calculation\cite{BC06}
to show that the sum of the $u_n$'s converges to the generalized extreme
value distributions $g_a(x)$, $f_{a,\mu}(x)$ or $w_{a,\mu}(x)$
defined in Eq.~(\ref{eqGk}) to (\ref{eqWk}).
This calculation has actually been done using a slightly more general
distribution $\Psi_{a,N'}(u_1,...,u_{N'})$
\be \label{PsiaM}
\Psi_{a,N'}(u_1,...,u_{N'}) = \frac{1}{Z_{N'}}\;
\Omega\left[ \tilde{F}\left(\sum_{n=1}^{N'} u_n\right)\right]
\prod_{n=1}^{N'} p\left(\sum_{i=n}^{N'} u_i\right),
\ee
where $\Omega(t)$ is an arbitrary positive function of $t>0$, and
where the normalization factor $Z_{N'}$ is given by
\be \label{eq-ZM}
Z_{N'} = \frac{1}{\Gamma(N')} \int_0^1 \dd t\, \Omega(t)\, (1-t)^{N'-1}.
\ee
In this case, the parameter $a>0$ characterizing the asymptotic distribution
is given by the small $t$ behavior of $\Omega(t)$, namely
\be
\Omega(t) \sim t^{a-1}, \qquad (t \rightarrow 0).
\ee
When $\Omega(t) = t^{a-1}$ for $0<t<1$, one has
\be
Z_{N'} = \frac{\Gamma(a)}{\Gamma(N'+a)},
\ee
so that one finds for $\Psi_{a,N'}(u_1,...,u_{N'})$ the straightforward
generalization of Eq.~(\ref{PhikN}).
If $\Omega(t)$ is only asymptotically a power law for $t \rightarrow 0$,
one does not recover
exactly the same form, but this does not affect the asymptotic distribution
which actually depends on the large $N'$ behavior of $Z_{N'}$, itself
dominated by the small $t$ behavior of $\Omega(t)$.

To sum up, the distribution of the sum $\sum_{n=1}^{N'} u_n$,
with the variables $u_n$ drawn
from the joint distribution $\Psi_{a,N'}(u_1,...,u_{N'})$, converges to
one of the three extreme value distributions, according to the asymptotic
large $z$ behavior of the function $p(z)$ appearing in Eq.~(\ref{PsiaM})
--and which also defines $\tilde{F}(z)$:

\begin{itemize}

\item If $p(z)$ decays at large $z$ faster than any power law, then
the asymptotic distribution of the sum is the generalized Gumbel
distribution $g_a(x)$.

\item If $p(z)$ decays as a power law $p(z) \sim z^{-(1+\mu)}$ ($\mu>0$)
when $z \rightarrow \infty$, the limit distribution is the generalized
Fr\'echet distribution $f_{a,\mu}(x)$.

\item If $p(z)=0$ for all $z>A$ and decays like $p(z) \sim (A-z)^{\mu-1}$
($\mu>0$) for $z \rightarrow A^{-}$, then the limit distribution is the
generalized Weibull distribution $w_{a,\mu}(x)$.

\end{itemize}

These results show that a large class of correlated variables,
with non-identical marginal distributions, lead as far as their sum
is concerned to the distributions found in extreme value statistics.
This fact may explain why these extreme value distributions, or at least
distributions qualitatively similar, are often found in correlated systems.

It is also worth mentioning that the distribution $\Psi_{a,N'}(u_1,...,u_{N'})$
may be symmetrized by summing over all possible permutations $\sigma$ of the
integers $1,\ldots,N'$:
\be
\Psi_{a,N'}^{\rm sym}(u_1,...,u_{N'}) = \frac{1}{N'!}
\sum_{\mathrm{perm} \;\sigma} \Psi_{a,N'}(u_{\sigma(1)},\ldots,u_{\sigma(N')}).
\ee
As the sum $\sum_{n=1}^{N'} u_n$ is invariant under permutation of
the terms, the distribution of the sum of random variables described
by $\Psi_{a,N'}^{\rm sym}(u_1,...,u_{N'})$ or by
$\Psi_{a,N'}(u_1,...,u_{N'})$ is the same.

To conclude on this point, let us mention that extreme value
distributions with a real index appeared recently in mathematics,
in the context of free probability.\cite{BenArous06} This topic is 
far beyond the scope of the present paper, but it could be interesting
to see if there is a contact point between the two problems.

To get a more intuitive feeling about the class of random variables
defined by the joint probability distribution $\Psi_{a,N'}(u_1,...,u_{N'})$,
it is interesting to focus on a simple example within the class.
Let us consider the case when $p(z)$ is a simple
exponential distribution, namely
\be
p(z) = \lambda\, e^{-\lambda z}, \qquad z>0,
\ee
leading for $\tilde{F}(z)$ to
\be
\tilde{F}(z) = \int_z^{\infty} \dd y \ p(y) = e^{-\lambda z}.
\ee
Then the distribution $\Psi_{a,N'}(u_1,...,u_{N'})$ reads
\be \label{PsiaM-exp}
\Psi_{a,N'}(u_1,...,u_{N'}) = \prod_{n=1}^{N'} \lambda (n+a-1)\,
e^{-\lambda(n+a-1)u_n}.
\ee
This is a generalization of the $1/f$-noise problem. Indeed, defining
$u_n=|c_n|^2$ in the $1/f$-noise model as defined in Eq.~(\ref{dist-cn}),
one precisely recovers the distribution (\ref{PsiaM-exp}), with
$\lambda=\kappa$ and $a=1$.

\subsubsection{Going backward from sums to extreme values}
\label{sect-back}

In the previous sections, we have seen how an extreme value problem
can be converted into a random sum problem, thanks to the
mapping relation (\ref{mapping}). A natural question would be
to know whether this relation also allows an inverse mapping to be
performed, that is to go from a problem of sum to a problem of
extreme value, which could lead for instance to Gaussian distributions
for extreme value problems, as already observed for instance in
problems of fluctuating fronts,\cite{Panja04} and in extreme values of
strongly correlated Gaussian sequences (see Theorems~\ref{thEVScorrel1} and 
\ref{thEVScorrel2}).
Although to our knowledge this possibility has not been explored yet
in the literature, such an inverse mapping can indeed
be performed, and we shall briefly sketch the argument in the following.

Consider a set of $N$ positive random variables $\{x_i\}_{i=1,\ldots,N}$
with a joint probability distribution $\Psi(x_1,\ldots,x_N)$. 
One can then generate a set of variables
$\{z_i\}$ such that
\be
z_i = \sum_{j=i}^N x_j
\ee
The joint distribution $J(z_1,\ldots,z_N)$ then reads
\bea
J(z_1,\ldots,z_N) &=& \int \prod_{i=1}^N \dd x_i\, \Psi(x_1,\ldots,x_N)
\prod_{i=1}^N \delta \left( z_i - \sum_{j=i}^N x_j \right)\\ \nonumber
&=& \Psi(z_1-z_2,z_2-z_3,\ldots,z_{N-1}-z_N,z_N)
\eea
However, the variables $\{z_i\}$ generated in this way form a increasing
sequence,
so that this distribution in some sense ``lacks randomness''.
This problem can easily be overcome by symmetrizing the distribution,
that is by summing over all possible permutations $\sigma$ over the
integers $1,\ldots,N$ (only one term in the sum is nonzero):
\be
J_{\mathrm{sym}}(z_1,\ldots,z_N) = \sum_{\mathrm{perm} \;\sigma}
J(z_{\sigma(1)},\ldots,z_{\sigma(N)}) \prod_{i=1}^{N-1}
\Theta(z_{\sigma(i)}-z_{\sigma(i+1)})
\ee
so that all variables play a symmetric role.
As an illustration, we consider two simples cases: the case when the variables
$x_i$ are i.i.d.~random variables, and the case when $\Psi(x_1,\ldots,x_N)$
is equal to $\Psi_{a,N}(x_1,\ldots,x_N)$ defined in Eq.~(\ref{PsiaM}).
We start with the case of i.i.d.~random variables, for which
\be
\Psi(x_1,\ldots,x_N) = \prod_{i=1}^N p(x_i)
\ee
If $p(x)$ is such that $\langle x^2 \rangle$ is finite, the maximum value
of the set of random variables $\{z_i\}$ drawn from
$J_{\mathrm{sym}}(z_1,\ldots,z_N)$ is, by construction, asymptotically
distributed according to a Gaussian distribution. This is actually
reminiscent of Theorems~\ref{thEVScorrel1} and \ref{thEVScorrel2}.
In the specific case when
$p(x)$ is a Gaussian distribution, then $J_{\mathrm{sym}}(z_1,\ldots,z_N)$
defines a Gaussian sequence. It would be interesting to know
whether this sequence fulfills the hypotheses underlying
Theorem~\ref{thEVScorrel1} or Theorem~\ref{thEVScorrel2}.
In contrast, if $\langle x^2 \rangle$ is infinite, the asymptotic distribution
of the maximum is a L\'evy-stable law.

As for the second case mentioned above, with
\be
\Psi(x_1,\ldots,x_N)=\Psi_{a,N}(x_1,\ldots,x_N),
\ee
defined in Eq.~(\ref{PsiaM}), one finds for $J_{\mathrm{sym}}(z_1,\ldots,z_N)$
\bea \nonumber
J_{\mathrm{sym}}(z_1,\ldots,z_N) &=& \sum_{\mathrm{perm}\;\sigma}\frac{1}{Z_N}
\,\Omega\left[\tilde{F}(z_{\sigma(1)})\right]
\left(\prod_{i=1}^N p(z_{\sigma(i)})\right) \prod_{i=1}^{N-1}
\Theta(z_{\sigma(i)}-z_{\sigma(i+1)})\\
&=& \frac{1}{Z_N} \left(\prod_{i=1}^N p(z_i)\right)\sum_{\mathrm{perm}\;\sigma}
\Omega\left[\tilde{F}(z_{\sigma(1)})\right]
\Theta(z_{\sigma(i)}-z_{\sigma(i+1)})
\eea
where the last step is obtained by relabeling $z_{\sigma(i)}$ in the product
$\prod_{i=1}^N p(z_{\sigma(i)})$. Then, given the constraints imposed
by the Heaviside functions, $z_{\sigma(1)}$ can be replaced by
$\max(z_1,\ldots,z_N)$, which is independent of the permutation $\sigma$.
The remaining sum over $\sigma$ of the product of Heaviside functions
is simply equal to one, because for a given set of values $\{z_i\}$,
only one permutation $\sigma$ satisfies all the constraints.
As a result, one finds
\be \label{eqJsym2}
J_{\mathrm{sym}}(z_1,\ldots,z_N) = \frac{1}{Z_N}\left(\prod_{i=1}^N
p(z_i)\right) \, \Omega\left[\tilde{F}(\max(z_1,\ldots,z_N))\right]
\ee
If $\Omega(t)$ is a constant, $J_{\mathrm{sym}}(z_1,\ldots,z_N)$ factorizes,
and one recovers a set of i.i.d.~random variables $\{z_i\}$.
This was expected since a constant $\Omega$ corresponds to
$a=1$ in the notations of Sect.~\ref{generalization}, that is to standard
extreme value statistics of i.i.d.~random variables.
The more general case $\Omega(t)=\Omega_0 t^{a-1}$ is also interesting;
we focus more specifically on the case when $p(z)$ decays faster than
any power law when $z \rightarrow \infty$, so that the asymptotic
distribution of $\max(z_1,\ldots,z_N)$ is the Gumbel $g_a(x)$.
In the large $a$ limit, we know that $g_a(x)$ converges to the normal
distribution. It is also expected from Eq.~(\ref{eqJsym2}) that
$J_{\mathrm{sym}}(z_1,\ldots,z_N)$ depends on $\{z_i\}$ mainly through
$\max(z_1,\ldots,z_N)$. Going beyond Eq.~(\ref{eqJsym2}),
one might think that if
$J_{\mathrm{sym}}(z_1,\ldots,z_N)=C_N\,\hat{J}(\max(z_1,\ldots,z_N))$,
where $\hat{J}$ is an arbitrary function, and $C_N$ is a normalization
constant, the asymptotic distribution of the maximum should exactly
be Gaussian.\footnote{This property can be verified explicitely in the
specific case when $\hat{J}(m)=e^{-\lambda x}$. In this case, the
distribution of the maximum is a Gamma distribution of index $N$, known
to converge to a Gaussian distribution when $N \rightarrow \infty$,
upon a correct rescaling.}

\subsection{Illustrations on one-dimensional physical models}
\label{phys-illust}

\subsubsection{A stochastic dissipative model}

Given that distributions rather similar to the generalized Gumbel
distribution have often been reported in the study of rather different
physical systems like critical magnetic models,\cite{BHP}
turbulent flow experiments,\cite{BHP,Pinton99} granular gases,\cite{Brey05}
or diverses nonequilibrium models,\cite{Bramwell01,Bramwell01b}
one could wonder whether some generic physical mechanism
is able to generate distributions of the microscopic variables
of the type shown in Eq.~(\ref{PsiaM}).
Indeed, the examples of turbulent flows and granular gases suggest
that dissipation
may play an important role in the emergence of non-Gaussian statistics
of the energy fluctuations. At equilibrium, equipartition of
energy implies that degrees of freedom with a quadratic energy all have
the same statistical properties, so that the total energy
generically has Gaussian fluctuations.

As a purpose of illustration, we consider a simple stochastic
model\cite{Bertin05,Bertin06} on a one-dimensional finite lattice,
in which energy is injected at one boundary, and dissipated in the bulk as
well as at the opposite boundary.
To be more specific, we label as $n=1,\ldots,N$
the sites of the lattice, and we define on each site $n$ a positive energy
variable $\ve_n$.
Injection of energy proceeds by adding an amount of energy $\nu$
on site $n=1$ with a probability rate $I(\nu)$.
Energy is transferred from site $n$ to site $n+1$ with a rate $\phi(\nu)$
(note that transport is fully biased), and energy may also
be dissipated at site $n$ with a rate $\Delta(\nu)$.
Note that the energy transferred from site $N$ is actually dissipated
since there is no site $N+1$.

A case of particular interest, which allows for a simple analytic solution
of the model, is when the different transition rates are
related in the following way\footnote{Note however that more general cases
have also been considered.\cite{Bertin06}}
\be
I(\nu) = e^{-\beta\nu} \, \phi(\nu), \qquad
\Delta(\nu) = (e^{\lambda\nu}-1)\, \phi(\nu).
\ee
The parameters $\beta$ and $\lambda$ respectively characterize the intensity
of the energy injection and of the energy dissipation with respect to
the energy transfer within the system.
From the master equation describing the evolution of the system,\cite{Bertin05}
one can show that the stationary joint distribution
$\mathcal{P}_{st}(\{\ve_n\})$ can be expressed as
\be
\mathcal{P}_{st}(\{\ve_n\}) = \prod_{n=1}^N (\lambda n+\beta)\,
e^{-(\lambda n+\beta) \ve_n}.
\ee
Clearly, this is the same distributions as $\Psi_{a,N'}(u_1,\ldots,u_{N'})$
given by Eq.~(\ref{PsiaM-exp}) provided that one identifies $\ve_n$ with $u_n$,
and $\beta$ with $(a-1)\lambda$.
Hence, it turns out that the fluctuations of the total energy
$E_N=\sum_{n=1}^N \ve_n$ are distributed --again up to a proper rescaling--
according to the generalized Gumbel distribution $g_a(x)$
in the limit $N \rightarrow \infty$.
The shape parameter $a$ is related to the physical parameters of the
model through:
\be
a = 1+\frac{\beta}{\lambda}.
\ee
This relation emphasizes the role of the dissipation in the statistics
of energy fluctuations, since when the dissipation is very small, that is
when $\lambda \rightarrow 0$, the shape parameter $a$ of the Gumbel
distribution goes to infinity, meaning that the distribution converges
to a Gaussian law.

It is interesting to note that another physical interpretation may be
given to the parameter $a$. To see this, we consider the lattice as
defined in Fourier space, and we map the sites $n$ of
the lattice described above onto wavenumbers $q_n=2\pi n/L$, where
$L$ is the system size (in real space).
In the spirit of the $1/f$-noise problem, one can also
interpret the energy $\ve_n$ as the squared Fourier amplitude $|c_n|^2$
associated to the wavenumber $q_n$. One then has
\be \label{cn2-tr}
\la |c_n|^2 \ra = \left( \frac{\lambda q_n L}{2\pi}+\beta \right)^{-1}.
\ee
On the other hand, the power spectrum $\la |c_n|^2 \ra$ is the Fourier
transform of the spatial correlation function of the system. Denoting as 
$\xi$ the correlation length, one can rewrite
Eq.~(\ref{cn2-tr}) in the form
\be
\la |c_n|^2 \ra = \frac{A}{q_n+\xi^{-1}},
\ee
with $A=2\pi/(\lambda L)$ and
\be
\xi = \frac{\lambda L}{2\pi\beta} = \frac{L}{2\pi(a-1)}.
\ee
This last result is particularly interesting. First, it shows that the
correlation length of the system is indeed proportional to the system
size $L$, as expected from the breaking of the central limit theorem.
Besides, it turns out that the ratio $\xi/L$ of the correlation length
to the system size is directly related to the ratio $\lambda/\beta$
comparing dissipation and injection, as well as to the
shape parameter $a$ of the Gumbel distribution. Interestingly, in the
$1/f$-noise problem, corresponding to the limit $a \rightarrow 1$, the ratio
$\xi/L$ becomes infinite, which may be interpreted as a highly correlated
limit. The model described here may then be considered as a model
of ``truncated $1/f$-noise''.

\subsubsection{A confined gas of classical independent particles}

In the last section, we gave an example of physical model leading
to the joint distribution (\ref{PsiaM}). However, this example deals
with the specific case of independent (though non-identically
distributed) variables. The joint distribution (\ref{PsiaM}) is actually
much more general, and it would be interesting to have a simple physical
example of correlated random variables following the distribution
(\ref{PsiaM}) in its general form.

\begin{figure}[!htb]
\begin{center}
\includegraphics[scale=0.33]{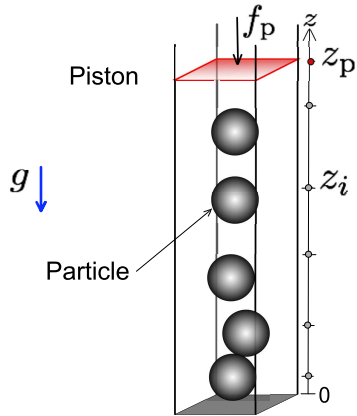}
\caption{Simple example: one-dimensional gas
of classical independent particles.}
\label{fig-gas1d}
\end{center}
\end{figure}

To this purpose, we consider the simple example of a one-dimensional gas
of classical independent particles.
Models of one-dimensional gases, like the Jepsen
gas,\cite{Frisch56,Jepsen65,Lebowitz67,Piasecki01,VandenBroeck02}
have proved very useful in the development of statistical physics,
due to their simplicity and to the possibility to test explicitely
some statistical approaches.
Here we consider a situation where the gas, subjected to an external
potential, is confined within a container closed by a piston
(see Fig.~\ref{fig-gas1d}), and is at
thermal equilibrium with a heat reservoir.\cite{BCH08}
The positions of the $N$ particles are denoted as $z_1,\ldots,z_N>0$,
and the piston is situated at $z_p$, with the obvious constraint
that $z_i<z_p$ for $i=1,\ldots,N$. Particles are subjected to an
external potential $U(z)$, such that $U(z) \rightarrow \infty$ when
$z \rightarrow \infty$. In addition, the piston is subjected to
a potential $U_p(z_p)$, that may be of the same type as that acting
on the particles, or of a different nature. For instance, the potential
$U_p(z_p)$ may describe a constant force $f_p=-|f_p|$ exerted by an external
operator, in which case $U_p(z_p)=|f_p| x_p$, or a harmonic potential
$U_p(z_p)=\frac{1}{2}kz_p^2$ created by spring.

At equilibrium at temperature $T$, the probability distribution of the
position of the particles and the piston reads
\be \label{eq-dist}
P_N(z_1,\ldots,z_N,z_p) = \frac{1}{\mathcal{Z}}\, e^{-\beta U_p(z_p)}
\prod_{i=1}^N e^{-\beta U(z_i)} \Theta(z_p-z_i),
\ee
with $\beta=1/k_BT$.
Similarly to what was done to map extreme values on sum in Eq.~(\ref{mapping}),
one can define the intervals $h_i$ between the ordered positions of the
particles, in such a way that the position of the piston, that is,
the volume of the system, can be expressed as
\be
z_p = \sum_{j=1}^{N+1} h_j.
\ee
Eq.~(\ref{eq-dist}) is then very similar to the distribution (\ref{PsiaM}),
up to a change of $N$ into $N+1$.
The identification is then made by choosing the function $p(z)$ as
\be
p(z) = \lambda\, e^{-\beta U(z)},
\ee
($\lambda$ is a constant such that $p(z)$ is normalized to one) and by
imposing that the function $\Omega(y)$ satisfies
\be \label{Omega-gas}
\Omega[F(z_p)] p(z_p) = e^{-\beta U_p(z_p)}.
\ee
A solution to this last equation can be found by noticing that $F(z_p)$
is a monotonous function of $z_p$, and is thus invertible.
We denote as $F^{-1}(y)$ the inverse function of $F(z_p)$.
Then the function $\Omega(y)$ can be defined from
\be
\Omega(y) = \frac{1}{\lambda} \exp\left[\beta U(F^{-1}(y))-\beta U_p(F^{-1}(y))\right],
\ee
a relation equivalent to Eq.~(\ref{Omega-gas}).
In the simple case when $U(z) = U_0\, z^{\alpha}$ and
$U_p(z_p) = U_0'\, z_p^{\alpha}$, with $U_0$, $U_0'$, $\alpha>0$,
the function $\Omega(y)$ behaves as a power law in the limit $y\rightarrow 0$,
\be
\Omega(y) \sim y^{a-1}, \qquad a=\frac{U_0'}{U_0},
\ee
up to logarithmic corrections.\cite{BCH08}
Then the distribution of the volume $z_p$ is, in the limit $N \rightarrow
\infty$, a generalized Gumbel distribution of parameter $a$.
More general situations can also be considered, leading either
to the Gaussian distribution, to the exponential distribution or to
the Fr\'echet or Weibull distributions.\cite{BCH08}

\subsection{An analog of the Gumbel distribution in two dimensions:
the 'BHP' distribution}

\subsubsection{Extending the 1/f-noise model to higher dimensions}

In the previous sections, we saw how, motivated by the study of the
$1/f$-noise model, extreme values statistics could be
reformulated as a problem of sums, allowing interesting extensions
to generalized extreme value distributions, as presented in
Sect.~\ref{generalization}. In particular, the class of random variables
defined by Eq.~(\ref{PsiaM}) turns out to be useful in the study of
some simple one-dimensional models, as seen in Sect.~\ref{phys-illust}.
In view of considering more realistic models, it is natural to try to
generalize the $1/f$-noise model to higher space dimensions.\footnote{Note
that interesting one-dimensional generalizations, known as
$1/f^{\alpha}$-noise, have also been studied.\cite{Racz02,Racz03}}
To this aim, one can consider $D$-dimensional sums of the form
\be
E = \sum_{\bm{q} \in \mathbb{Z}_D} \mathcal{G}^{-1}(\bm{q})
\theta_{\bm{q}}\theta_{-\bm{q}},
\ee
where $\mathbb{Z}_D$ is the $D$-dimensional Brillouin zone, \textit{i.e.},
the set of $D$-dimensional vectors $\bm{q}=(2\pi n_1/L,\cdots,2\pi n_D/L)$,
where $0 \le n_k \le L-1$ ($k=1,\ldots,D$), and
$\theta_{\bm{q}}$ is the Fourier coefficient associated with mode $\bm{q}$.
Note that for convenience, we use here notations coming from field theory,
but one could equally use the summation on integers
as done in Sect.~(\ref{noise1}).
For simplicity, we search for a model that fulfills rotational invariance
(at least at large scale), so that we choose the inverse propagator
$\mathcal{G}^{-1}(\mathbf{q})$ to depend only on the modulus $q=|\bm{q}|$.
An essential property of the $1/f$-noise model is the logarithmic behavior
of the average sum with the number $N$ of terms. One thus wishes to choose the
most simple function $\mathcal{G}^{-1}(q)$ that fulfills this property
in arbitrary dimension. In the continuous limit, one has
\be
\langle E_L \rangle \approx \int_{2\pi/L}^{2\pi}
\dd q\ \rho(q)\, \mathcal{G}(q)
\ee
with $N=L^D$, and where $\rho(q)$ is the coarse-grained density of modes
on the reciprocal lattice.
One has $\rho(q) \sim q^{D-1}$, so that $E_L \sim \ln L$
on condition that $\mathcal{G}(q) \sim q^{-D}$.
We thus simply take $\mathcal{G}^{-1}(q)=q^D$.

The two-dimensional case deserved a lot of attention in the recent
literature, since the propagator $\mathcal{G}(q)=q^{-2}$ is
very frequent in physical systems.
It describes for instance the magnetization of the
two-dimensional XY-model\cite{Archambault97} in the low-temperature regime,
as well as the roughness of fluctuating interfaces in the two-dimensional
Edwards-Wilkinson model.\cite{EW82,Racz94,Bramwell01b}
The corresponding distribution is often called "BHP distribution" after
Bramwell, Holdsworth and Pinton who first reported it in turbulent
flows and in the XY-model.\cite{BHP}

\begin{figure}[!htb]
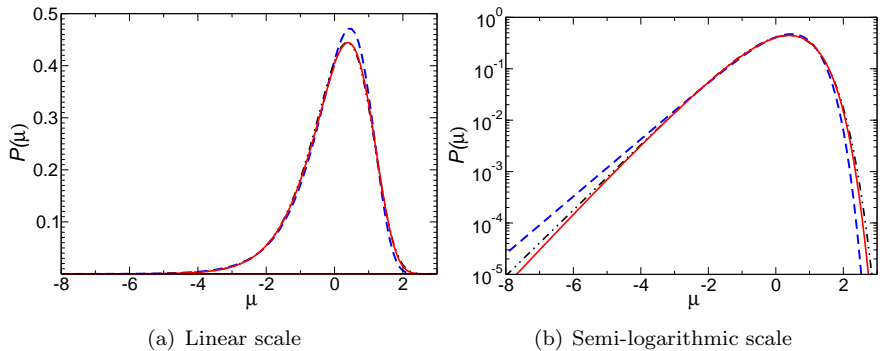

\begin{center}
\subfigure[Linear scale]{\includegraphics[width=0.45\linewidth]{gumbbhp-lin.eps}}
\subfigure[Semi-logarithmic scale]{\includegraphics[width=0.45\linewidth]{gumbbhp-log.eps}}
\caption{Comparison between standard Gumbel distribution ($a=1$, dash),
generalized Gumbel distribution with $a=1.57$ (dot dash),
and BHP distribution (plain).}
\label{Gumbel-BHP}
\end{center}
\end{figure}

In the context of the two-dimensional XY-model, the global quantity of interest is the total magnetization $M$ of the sample of $N$ spins, defined in terms of independent Fourier coefficients $\theta_{\bm{q}}$ by
\be \label{magnetisation}
M=1-\frac{1}{N} \sum_{\bm{q} \in \mathbb{Z}_D} \theta_{\bm{q}}\theta_{-\bm{q}}.
\ee
Bramwell and co-workers obtained an expression for the
Fourier transform of the BHP distribution,\cite{Bramwell01b} under some
hypotheses that have been questioned and clarified
later on.\cite{Banks05,Mack05}
Although the BHP distribution is not of Gumbel type,
it has been shown that it could be
quite well approximated (for instance by matching the first four cumulants)
by a generalized Gumbel distribution with parameter
$a \approx 1.57$, in the experimentally or numerically accessible
window\cite{Bramwell01,Bramwell01b} (see Fig.~\ref{Gumbel-BHP}).
This similarity with the (generalized) Gumbel
distribution is confirmed by the asymptotic expressions of the
tails:\cite{Bramwell01b}
\bea
P_{\mathrm{BHP}}(x) &\sim& x\, e^{\lambda_1 x}, \qquad x\rightarrow -\infty\\
P_{\mathrm{BHP}}(x) &\sim& \exp(-\eta \, e^{\lambda_2 x}+\lambda_2 x)
\qquad x\rightarrow +\infty,
\eea
with some positive constants $\eta$, $\lambda_1$ and $\lambda_2$.
Note that because of the minus sign in the definition of the magnetization
(\ref{magnetisation}), the asymmetry is opposed to the one of the standard
extreme value distributions presented in section (\ref{iidext1}).
These asymptotic relations are interesting, since they show at the same time
that the tails are qualitatively similar to that of the Gumbel distribution,
but that there exists quantitative difference, like
the algebraic prefactor in the positive tail.

\subsubsection{BHP distribution as the leading contribution in the moderately
correlated regime}

It has been early noticed that the BHP distribution also appears, within
experimental errors, in a context completely different from magnetic
systems, namely the fluctuations of injected power in a three-dimensional
driven turbulent flow.\cite{BHP} This amazing observation
triggered many other studies where the BHP distribution
has been found to be a good approximate description of experimentally
or numerically measured distributions.\cite{Pinton99,Noullez02,Tothkatona03,Pennetta04,Vanmil05,Brey05}
The possible reasons for this apparent ubiquity have been quite debated
in the last decade,\cite{Jensen01,Watkins02,Bramwell02,Bertin05,BC06}
particularly due to a tentative link with extreme
events.

\begin{figure}[t]
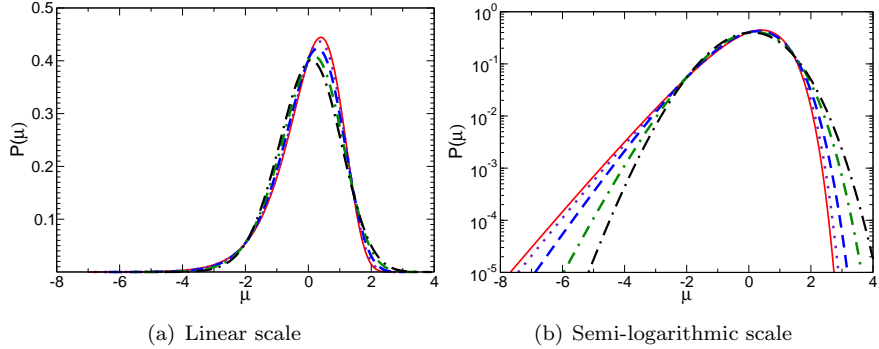

\begin{center}
\subfigure[\label{massivexy-lin}Linear scale]{\includegraphics[width=0.45\linewidth]{massiveXY-lin.eps}}
\subfigure[\label{massivexy-log} Semi-logarithmic scale]{\includegraphics[width=0.45\linewidth]{massiveXY-log.eps}}
\caption{\label{massivexy} Distributions obtained for the Gaussian model
using propagator (\ref{gaussianprop}), for different masses: $\phi_0=0$
('BHP' distribution, plain), $\phi_0=0.1$ (dot), $\phi_0=1$ (dash),
$\phi_0=2$ (dot dash) and $\phi_0=5$ (long dash). Note the convergence
towards the normal distribution when the masse increases.}
\end{center}
\end{figure}

We now give a generic argument, inspired by the study of the two-dimensional
Ising model,\cite{Clusel04,Clusel06} in order to clarify this issue.
Consider a system fluctuating around a local minimum of potential energy;
this could be a system at equilibrium, or a particular case of
out-of-equilibrium system. The distribution of the fluctuations of the
global quantity is described by a (pseudo-)Hamiltonian,
or 'action' in a field-theoretic language, which could be quite complex.
In order to obtain a Gaussian effective action, the simplest idea is to
perform a perturbative expansion. From the probability viewpoint, this is
similar to group the random variables by blocks of linear size
of the order of the correlation length.
It is however well-known from the theory of critical phenomena that the
resulting effective action does not describe correctly all the system
properties. The breakdown of this expansion is quantified by the
Ginzburg criterion.\cite{Goldenfeld92} We define the ratio
\be
\label{criterion}
R_{\xi}=\ff{\int_{S(\xi)} C(\bm{r})\dd
\bm{r}}{\int_{S(\xi)} \langle m(\bm{r})\rangle ^2 \dd \bm{r}},
\ee
where $\xi$ is the correlation length, $C(\bm{r})$ the two-point correlation
function and $m(\bm{r})$ the local order parameter (to be interpreted
as the local magnetization in the Ising model); the integrals are over
a sphere of radius $\xi$ in $D$ dimensions.
A mean-field theory captures the physical behavior of the system
if $R_{\xi} \ll 1$.\cite{Goldenfeld92}
In the present context, we do not look for an effective description
of all the properties
of the system at the scale of the correlation length, but we only want to
describe the behavior of its large scale fluctuations. A natural extension
of the previous criterion in order to check the validity of a perturbative
approach at the integral scale is then\cite{Clusel06}
\be \label{criteriongen}
R_L=\ff{\int_{S(L)} C(\bm{r})\dd
\bm{r}}{\int_{S(L)} \langle m(\bm{r})\rangle ^2 \dd \bm{r}} =
A \left(\frac{\xi}{L}\right)^D \ll 1,
\ee
where $A$ is an order unity constant.
This criterion is actually much less restrictive than the original
criterion. This means that the large scale behavior of a system could
be studied by a perturbative expansion much more easily than the small scale
behavior. In particular, for the two-dimensional Ising model,
it is possible to perform
such an expansion while the (finite size) system is already in the
critical region:\cite{Clusel06} starting in the low temperature regime,
we approach the critical point from below, so that $\xi$ increases.
We then increase the system size while keeping the ratio $\xi / L$ constant,
but much less than $1$.
Doing so the correlation length diverges, confirming the
critical behavior of the system considered. But it diverges with a small
amplitude which allows the large scale behavior to be captured by a
perturbation theory.
The action can then be decomposed in the following way:
\be
\mathcal{S}=\frac{1}{2}\sum_{\bm{q} \neq 0}
\mathcal{G}^{-1}(\bm{q},\phi_0)
\theta_{\bm{q}}\theta_{-\bm{q}}+\mathcal{S}_0(\phi_0),
\ee
where 
\be\label{gaussianprop}
\mathcal{G}^{-1}(\bm{q},\phi_0)=\phi_0+q_x^2+q_y^2,
\ee
and $\phi_0$ is the position of the energy minimum.
A detailed study of those fluctuations,\cite{Clusel06,CluselTh}
shows that the first part of the 
action leads to fluctuations similar to that of the two-dimensional XY-model
at low temperature, with a cut-off related to $\phi_0$, playing a role
similar to that of a magnetic field in the XY-model.\cite{Portelli01}
This contribution does not depend on the universality
class (in the sense of critical phenomena) of the system,
and therefore could be seen as "super-universal"; it depends only
on the space dimension $D$.
If $\phi_0$ can be varied by adjusting the external control
parameters (like temperature or magnetic field), reducing its value
close to zero should make the distribution closer to the BHP one,
as illustrated on Fig.~\ref{massivexy}.
However, doing so, the effect of higher order terms (in particular quartic
terms) in the action may become significant, leading to 
corrections with respect to the BHP distribution.
These corrections correspond to the fact that the system could leave 
the local minima and explore the whole phase space; they are
characteristic of the system considered, and can be formally computed
using instantons.\cite{Leung82}

\begin{figure}[t]
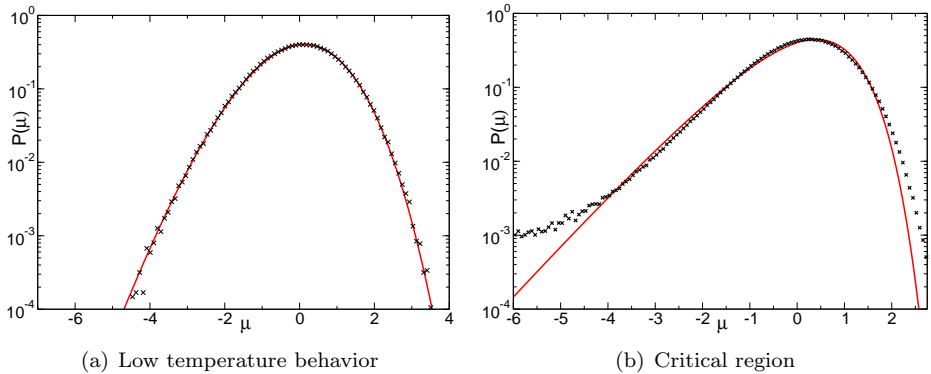

\begin{center}
\subfigure[Low temperature behavior]{
\includegraphics[width=0.47 \linewidth]{pdfferro.eps}
}
\subfigure[Critical region]{
\includegraphics[width=0.47 \linewidth]{pdfTet.eps}
} 
\caption{(a) Ising model at $T=1.7J$ for $N=64^2$
(cross), and Gaussian model with mass $\phi_0=0.3$ (plain).
(b) Ising model at $T=2.13J$ for $N=64^2$ (cross),
and BHP distribution (plain), corresponding to a 2D gaussian model with zero
mass.}
\label{pdfH1}
\end{center}
\end{figure}

The contribution of these corrections in the probability distribution
is actually rather easy to identify.
By construction the large negative fluctuations induced by
$\mathcal{G}^{-1}(\bm{q},\phi_0)$ are closer
to the Gaussian distribution than to the BHP distribution, given by
$\mathcal{G}^{-1}(\bm{q},0)$. In particular the quasi-exponential
tail should be below the one of the BHP distribution,
as seen on Fig.~\ref{pdfH1}, which compares numerical simulations
of the two-dimensional Ising model and of the two-dimensional Gaussian model.
At low temperature, Fig.~\ref{pdfH1}(a), the two distributions overlap,
showing that non-linear contributions are negligible.
While we approach the critical point from below, Fig.~\ref{pdfH1}(b),
the weight of such corrections increases, so that it becomes impossible
to neglect them to describe the large fluctuations.
These observations lead to a practical criterion to say whether the
large scale behavior of the system could be studied through a perturbative
approach.

Altogether, the above approach --which can be generalized in a straightforward
way to $D$-dimensional systems-- yields a rationale for the frequent
appearance of distributions close to the BHP distribution in two-dimensional
systems. This comes from the rather general
relevance of Gaussian models with propagator $\mathcal{G}^{-1}(\bm{q},0)$
in the description of fluctuations around a local minimum of energy.
Note however that, quite importantly, this does not explain the
observation of the BHP distribution in turbulent flows: assuming that
a Gaussian action could be a reasonable description of such a flow
(which would remain to be justified), one would expect a three-dimensional
Gaussian model to be relevant in this case. This might suggest that
a dimensional reduction is at play in out-of-equilibrium systems.
Alternatively, it may be possible that a Gaussian action is not a relevant
description in this case, and that the reason for the appearance of an
approximate BHP distribution in this context relies on completely different
grounds.
In any case, understanding the statistical behavior of
driven systems, and the possible relationships with (critical) equilibrium
systems, remains one of the challenges of nonequilibrium statistical physics.

\section{Conclusion}

In the present review article, we tried to adopt a somehow original
perspective by presenting essentially on the same footing different
types of convergence theorems, namely those for sums of random variables
and those for extremes values. The motivations for this was to highlight
the similarities, differences and relationships between these two
a priori unrelated fields of probability and statistics.
As this review paper is mainly aimed at a physicist readership, we attempted
to incorporate both the mathematical rigor when stating the theorems,
and more intuitive physicist's approaches when going farther than
precise mathematical results would allow us to go. We also tried to give
many physical examples of applications, in order to make contact
between formal results and the real world, or at least some simplified
pictures of the latter.

Beside recalling the standard mathematical convergence theorems, one of the
central points of the present paper is the mapping between extreme values
and sums of random variables, which precisely allows one to bridge the gap
between the two fields. Although this mapping is very simple, it has
non trivial consequences, in that it allows to map the convergence
theorems known in one field to the other field. In particular,
generalized extreme value distributions can be recovered for sums
of specific classes of non-i.i.d.~variables, and Gaussian distributions
are found for the extreme value of some classes of correlated random variables.
Another bridge between the two fields is also
the dominant effect of the largest terms within sums of broadly
distributed variables. Still, in this case, the relation is very simple
(and more direct), and relies on the typical magnitude of the extreme values
rather than on the specific properties of their fluctuations.
Moreover, the sum does not have exactly the same distribution as
its largest term, only the scaling properties with the number of terms is
the same.

Clearly, statistical physics remains a fertile testing ground for
probability concepts, and both fields benefit from this mutual interaction.
In this spirit, let us mention a recent example motivated by the physics
of disordered systems, and more specifically the Random Energy
Model.\cite{Derrida81,Derrida81b}
In this model, the partition function, which is a random variable due to the
presence of random disorder, is a sum of independent random contributions
(the Boltzmann-Gibbs weights associated to microscopic configurations).
One of the interests of this problem comes from the fact that
the statistics of each
term depends on the total number of terms, both being related to the system
size. Hence standard convergence theorems cannot be applied, and
this induces a non-trivial behavior, which is at the origin
of the glass transition in this model.
Recent progresses have been made in this specific field,\cite{BenArous05}
and one may hope that such results will trigger new researches
both in mathematics and physics.

\section*{Acknowledgments}

The authors are grateful to S.~T. Banks, S.~T. Bramwell, M. Droz,
J.-Y. Fortin and P.~C.~W. Holdsworth for interesting comments and
helpful discussions.


\section*{References}

\begin{thebibliography}{99}

\bibitem{Mon73}
E.~W. Montroll and H. Scher, \textit{J. Stat. Phys.} \textbf{9}, 101 (1973).

\bibitem{Shl74}
M.~F. Shlesinger, \textit{J. Stat. Phys.} \textbf{10}, 421 (1974).

\bibitem{Mon75}
H. Scher and E.~W. Montroll, \textit{Phys. Rev.} \textbf{B12}, 2455 (1975).

\bibitem{Shl88}
M.~F. Shlesinger, \textit{Ann. Rev. Phys. Chem.} \textbf{39}, 269 (1988).

\bibitem{BG}
J.-P. Bouchaud and A. Georges, \textit{Phys. Rep.} \textbf{195}, 127 (1990).

\bibitem{Bouchaud92}
J.-P. Bouchaud, \textit{J. Phys. I} (France) \textbf{2}, 1705 (1992).

\bibitem{Bouchaud95}
J.-P. Bouchaud and D.~S. Dean, \textit{J. Phys. I} (France) \textbf{5}, 265
(1995).

\bibitem{BBE94}
F. Bardou, J.-P. Bouchaud, O. Emile, A. Aspect, and C. Cohen-Tannoudji,
\textit{Phys. Rev. Lett.} \textbf{72}, 203 (1994).

\bibitem{BBA02}
F. Bardou, J.-P. Bouchaud, A. Aspect, and C. Cohen-Tannoudji,
\textit{L\'evy Statistics and Laser Cooling} (Cambridge University Press,
Cambridge, 2002).

\bibitem{SZK93}
M.~F. Shlesinger, G.~M. Zaslavsky, and J. Klafter, \textit{Nature}
\textbf{363}, 31
(1993).

\bibitem{SWS93}
T.~H. Solomon, E.~R. Weeks, and H.~L. Swinney, \textit{Phys. Rev. Lett.}
\textbf{71}, 3975 (1993).

\bibitem{Min96}
I.~A. Min, I. Mezic, A. Leonard, \textit{Phys. Fluids} \textbf{8}, 1169 (1996).

\bibitem{Brokmann}
X. Brokmann, J.-P. Hermier, G. Messin, P. Desbiolles, J.-P. Bouchaud, and
M. Dahan, \textit{Phys. Rev. Lett.} \textbf{90}, 120601 (2003).

\bibitem{Bruce81}
A.~D. Bruce, \textit{J. Phys.} \textbf{C14}, 3667 (1981).

\bibitem{Binder81}
K. Binder, \textit{Z. Phys.} \textbf{B43}, 119 (1981).

\bibitem{Binder92}
K. Binder, \textit{Computational Methods in Field Theory}
(Springer, Berlin, 1992).

\bibitem{Zheng01}
B. Zheng and S. Trimper, \textit{Phys. Rev. Lett.} \textbf{87}, 188901 (2001).

\bibitem{Zheng03}
B. Zheng, \textit{Phys. Rev.}\textbf{E67}, 026114 (2003).

\bibitem{Plischke94}
M. Plischke, Z. R\'acz, and R.~K.~P. Zia, \textit{Phys. Rev.}\textbf{E50},
3589 (1994).

\bibitem{Racz94}
Z. R\'acz and M. Plischke, \textit{Phys. Rev.}\textbf{E50}, 3530 (1994).

\bibitem{Verma00}
M.~K. Verma, \textit{Physica} \textbf{A277}, 359 (2000).

\bibitem{Marinari02}
E. Marinari, A. Pagnani, G. Parisi, and Z. R\'acz, \textit{Phys. Rev.}
\textbf{E65}, 026136 (2002).

\bibitem{Rosso03}
A. Rosso, W. Krauth, P. Le Doussal, J. Vannimenus, and K.~J. Wiese,
\textit{Phys. Rev.}\textbf{E68}, 036128 (2003).

\bibitem{Moulinet04}
S. Moulinet, A. Rosso, W. Krauth, and E. Rolley, \textit{Phys. Rev.}
\textbf{E69}, 035103 (2004).

\bibitem{Frisch95}
U. Frisch, \textit{Turbulence} (Cambridge Univ. Press, Cambridge, 1995).

\bibitem{Peinke97}
R. Friedrich and J. Peinke, \textit{Phys. Rev. Lett.} \textbf{78}, 863 (1997).

\bibitem{Carreras99}
B.~A. Carreras \textit{et al.}, \textit{Phys. Rev. Lett.} \textbf{83}, 3653
(1999).

\bibitem{Portelli03}
B. Portelli, P.~C.~W. Holdsworth, and J.-F. Pinton, \textit{Phys. Rev. Lett.}
\textbf{90}, 104501 (2003).

\bibitem{Chevillard05}
L. Chevillard, S.~G. Roux, E. Lev\`eque, N. Mordant, J.-F. Pinton, and
A. Arn\'eodo, \textit{Phys. Rev. Lett.} \textbf{95}, 064501 (2005).

\bibitem{Aumaitre01}
S. Auma\^{\i}tre, S. Fauve, S. McNamara, and P. Poggi, \textit{Eur. Phys. J.}
\textbf{B19}, 449 (2001).

\bibitem{Farago02}
J. Farago, \textit{J. Stat. Phys.} \textbf{107}, 781 (2002).

\bibitem{Droz03}
F. Coppex, M. Droz, J. Piasecki, and E. Trizac, \textit{Physica} \textbf{A329},
114 (2003).

\bibitem{Visco06}
P. Visco, A. Puglisi, A. Barrat, F. van Wijland, and E. Trizac,
\textit{Eur. Phys. J.} \textbf{B51}, 377 (2006).

\bibitem{Weill05}
R. Weill, A. Rosen, A. Gordon, O. Gat, and B. Fischer,
\textit{Phys. Rev. Lett.} \textbf{95}, 013903 (2005).

\bibitem{Pilgram03}
S. Pilgram, A.~N. Jordan, E.~V. Sukhorukov, and M. B\"uttiker,
\textit{Phys. Rev. Lett.} \textbf{90}, 206801 (2003).

\bibitem{Reulet03}
B. Reulet, J. Senzier, and D.~E. Prober, \textit{Phys. Rev. Lett.} \textbf{91},
196601 (2003).

\bibitem{BHP}
S.~T. Bramwell, P.~C.~W. Holdsworth and J.-F. Pinton, \textit{Nature}
\textbf{396}, 552 (1998).

\bibitem{Pinton99}
J.-F. Pinton, P.~C.~W. Holdsworth and R. Labb\'e, \textit{Phys. Rev.}
\textbf{E60}, R2452 (1999).

\bibitem{Bramwell01}
S.~T. Bramwell \textit{et al.} \textit{Phys. Rev. Lett.} \textbf{87}, 188902
(2001).

\bibitem{Bramwell01b}
S.~T. Bramwell \textit{et al.}, \textit{Phys. Rev.} \textbf{E63} 041106 (2001).

\bibitem{Noullez02}
A. Noullez and J.-F. Pinton, \textit{Eur. Phys. J.} \textbf{B28}, 231 (2002).

\bibitem{Holdsworth02}
S.~T. Bramwell, T. Fennell, P.~C.~W. Holdsworth, and B. Portelli,
\textit{Europhys. Lett.} \textbf{57}, 310 (2002).

\bibitem{Tothkatona03}
T. T\'oth-Katona and J. Gleeson, \textit{Phys. Rev. Lett.} \textbf{91}, 264501
(2003).

\bibitem{Fedorenko03}
A.~A. Fedorenko and S. Stepanow, \textit{Phys. Rev.} \textbf{E68},
056115 (2003).

\bibitem{Pennetta04}
C. Pennetta \textit{et al.}, \textit{Semicond. Sci. Technol.}, \textbf{19},
S164 (2004).

\bibitem{Chamon04}
C. Chamon \textit{et al.}, \textit{J. Chem. Phys.} \textbf{121}, 10120 (2004).

\bibitem{Wijland04}
F. van Wijland, \textit{Physica} \textbf{A332}, 360 (2004).

\bibitem{Vanmil05}
B.~Ph. van Milligen \textit{et al.}, \textit{Phys. of Plasmas} \textbf{12},
052507 (2005).

\bibitem{Brey05}
J.~J. Brey, M.~I. Garcia de Soria, P. Maynar and M.~J. Ruiz-Montero,
\textit{Phys. Rev. Lett.} \textbf{94}, 098001 (2005).

\bibitem{Duri05}
A. Duri, H. Bissig, V. Trappe and L. Cipelletti, \textit{Phys. Rev. E}
\textbf{72}, 051401 (2005).

\bibitem{Varotsos05}
P.~A. Varotsos, N.~V. Sarlis, H.~K. Tanaka and E.~S. Skordas,
\textit{Phys. Rev.} \textbf{E72}, 041103 (2005).

\bibitem{Feynman}
R.~P. Feynman, \textit{The Character of Physical Laws}, (MIT Press, 1967).

\bibitem{Feller}
W. Feller, \textit{An Introduction to Probability Theory and its Applications}
(Wiley, New York, 1966), Vol. I.

\bibitem{Kolmogorov}
B.~V. Gnedenko and A.~N. Kolmogorov, \textit{Limit Distributions for Sums of
Independent Random Variables} (Addisson-Wesley, 1954).

\bibitem{Levy}
P. L\'evy, \textit{Th\'eorie de l'Addition de Variables
Al\'eatoires} (Gauthier-Villard, Paris, 1954; J. Gabay, 2003).

\bibitem{Feller2}
W. Feller, \textit{An Introduction to Probability Theory and its Applications}
(Wiley, New York, 1966), Vol. II.

\bibitem{Petrov75}
V.~V. Petrov, \textit{Sums of Independent Random Variables} (Springer-Verlag,
Berlin, 1975).

\bibitem{Petrov95}
V.~V. Petrov, \textit{Limit Theorems of Probability Theory, Sequences of
Independent Random Variables} (Oxford University Press, Oxford, 1995). 

\bibitem{Gnedenko}
B. Gnedenko, \textit{Ann. Math.} \textbf{44}, 423 (1943).

\bibitem{Galambos}
J. Galambos, \textit{The Asymptotic Theory of Extreme Order 
Statistics} (Wiley, New York, 1987).

\bibitem{Gumbel}
E.~J. Gumbel, \textit{Statistics of Extremes} (Columbia
University Press, New York, 1958; Dover Publication, 2004).

\bibitem{Embrechts97}
P. Embrechts, C. Kl\"uppelberg, and T. Mikosch,
\textit{Modelling Extremal Events} (Springer, Berlin, 1997).

\bibitem{Reiss01}
R. Reiss and M. Thomas, \textit{Statistical Analysis of Extreme Values}
(Birkh\"auser, Basel, 2001).

\bibitem{Falk}
M. Falk, J. H\"usler and R.-D Reiss, \textit{Laws of Small Numbers: Extremes
and Rare Events} (Birkh\"auser, Basel, 2004).

\bibitem{Coles}
S. Coles, \textit{An Introduction to Statistical Modeling of Extreme Values},
Springer Series in Statistics (Springer-Verlag London, Berlin,
Heidelberg 2001).

\bibitem{Racz07}
G. Gyorgyi, N.~R. Moloney, K. Ozogany, Z. R\'acz, arXiv:0712.3993.

\bibitem{Jona01}
G. Jona-Lasinio, \textit{Physics Reports} \textbf{352}, 439-458 (2001).

\bibitem{Poincare}
H. Poincar\'e, \textit{Thermodynamique} (Georges Carr\'e, Paris, 1892),
cited by J.~H. Gaddum, Nature \textbf{156}, 463 (1945).

\bibitem{ZuK94}
G. Zumofen and J. Klafter, \textit{Chem. Phys. Lett.} \textbf{219}, 303 (1994).

\bibitem{Barkai}
E. Barkai, A.~V. Naumov, Y.~G. Vainer, M. Bauer, and I. Kador,
\textit{Phys. Rev. Lett.} \textbf{91}, 075502 (2003).

\bibitem{OBL90}
A. Ott, J.-P. Bouchaud, D. Langevin, and W. Urbach, \textit{Phys. Rev. Lett.}
\textbf{65}, 2201 (1990).

\bibitem{MEZ96}
S. Marksteiner, K. Ellinger, and P. Zoller, \textit{Phys. Rev.} \textbf{A53},
3409 (1996).

\bibitem{KSW97}
H. Katori, S. Schlipf, and H. Walther, \textit{Phys. Rev. Lett.} \textbf{79},
2221 (1997).

\bibitem{Monthus96}
C. Monthus and J.-P. Bouchaud, \textit{J. Phys.} \textbf{A29}, 3847 (1996).

\bibitem{Bertin02}
E. Bertin and J.-P. Bouchaud, \textit{J. Phys.} \textbf{A35}, 3039 (2002).

\bibitem{Rinn00}
B. Rinn, P. Maass, and J.-P. Bouchaud, \textit{Phys. Rev. Lett.} \textbf{84},
5403 (2000).

\bibitem{Bertin03}
E.~M. Bertin and J.-P. Bouchaud, \textit{Phys. Rev.} \textbf{E67}, 026128
(2003).

\bibitem{Bertin03b}
E.~M. Bertin and J.-P. Bouchaud, \textit{Phys. Rev.} \textbf{E67}, 065105(R)
(2003).

\bibitem{Monthus03}
C. Monthus, \textit{Phys. Rev.} \textbf{E68}, 036114 (2003).

\bibitem{Bardou05}
E. Bertin and F. Bardou, \textit{Am. J. Phys.} \textbf{76}, 630 (2008).

\bibitem{Blinnikov}
S. Blinnikov and R. Moessner, \textit{Astron. Astrophys. Suppl. Ser.}
\textbf{130}, 193 (1998).

\bibitem{Dacosta00}
V. da Costa, Y. Henry, F. Bardou, M. Romeo, and K. Ounadjela,
\textit{Eur. Phys. J.} \textbf{B13}, 297 (2000).

\bibitem{Bardou97}
F. Bardou, \textit{Europhys. Lett.} \textbf{39}, 239 (1997).

\bibitem{Romeo03}
M. Romeo, V. da Costa and F. Bardou, \textit{Eur. Phys. J.} \textbf{B32}, 513
(2003).

\bibitem{BouchMez97}
J.-P. Bouchaud and M. M\'ezard, \textit{J. Phys.} \textbf{A30}, 7997 (1997).

\bibitem{Baldassarri02}
A. Baldassarri, A. Gabrielli and B. Sapoval,
\textit{Europhys. Lett.} \textbf{59}, 232 (2002).

\bibitem{Katz02}
R.~W. Katz, M.~B. Parlangi and P. Naveau, \textit{Advances in Water
Ressources} \textbf{25}, 1287 (2002).

\bibitem{Sornette96}
D. Sornette, L. Knopoff, Y. Kagan and C. Vannest, \textit{J. Geophys. Res.}
\textbf{101}, 13883 (1996).

\bibitem{Longin00}
F. Longin, \textit{J. Bank. Finance} \textbf{24}, 1097 (2000).

\bibitem{Potters}
J.-P. Bouchaud and M. Potters, \textit{Theory of Fincancial Risk and Derivative
Pricing} (Cambridge University Press, Cambridge, 2003), 2nd edition.

\bibitem{Mises}
R. von Mises, reprinted in \textit{Selected Papers II}, Amer. Math. Soc.,
Providence, R.I. (1954).

\bibitem{Antal01}
T. Antal, M. Droz, G. Gy\"orgyi and Z. R\'acz, \textit{Phys. Rev. Lett.}
\textbf{87}, 240601 (2001).

\bibitem{BC06}
E. Bertin and M. Clusel, \textit{J. Phys.} \textbf{A39}, 7607 (2006).

\bibitem{GR}
I.~S. Gradshteyn and I.~M. Ryzhik, \textit{Table of Integrals, Series,
and Products} (Academic Press, London, 5th edition, 1994).

\bibitem{Schehr06}
G. Schehr and S.~N. Majumdar, \textit{Phys. Rev.} \textbf{E73}, 056103 (2006).

\bibitem{Comtet07}
A. Comtet, P. Leboeuf, and S.~N. Majumdar, \textit{Phys. Rev. Lett.}
\textbf{98}, 070404 (2007)

\bibitem{Billingsley}
P. Billingsley, \textit{Proc. Am. Math. Soc.} \textbf{12}, 788 (1961).

\bibitem{Ibragimov}
I.~A. Ibragimov, \textit{Theo. Proba. Appl.} \textbf{8}, 83 (1963).

\bibitem{Sun65}
T.~C. Sun, \textit{J. Math. and Mech.} \textbf{14}, 71 (1965).

\bibitem{Taqqu77}
M.~S. Taqqu, \textit{Z. Wahrsch. Verw. Gebiete} \textbf{40}, 203 (1977).

\bibitem{Dobrushin79}
R.~L. Dobrushin and P. Major, \textit{Z. Wahrsch. Verw. Gebiete} \textbf{50},
27 (1979).

\bibitem{Taqqu79}
M.~S. Taqqu, \textit{Z. Wahrsch. Verw. Gebiete} \textbf{50}, 53 (1979).

\bibitem{Rosenblatt81}
M. Rosenblatt, \textit{Z. Wahrsch. Verw. Gebiete} \textbf{55}, 123 (1981).

\bibitem{Major81}
P. Major, \textit{Z. Wahrsch. Verw. Gebiete} \textbf{57}, 129 (1981).

\bibitem{Breuer83}
P. Breuer and P. Major, \textit{J. Multiv. Anal.} \textbf{13}, 425 (1983).

\bibitem{Baldovin07}
F. Baldovin and A.~L. Stella, \textit{Phys. Rev.} \textbf{E75}, 020101(R)
(2007).

\bibitem{Tsallis06a}
S. Umarov, C. Tsallis, and S. Steinberg, cond-mat/0603593.

\bibitem{Tsallis06b}
S. Umarov, C. Tsallis, M. Gell-Mann, and S. Steinberg, cond-mat/0606038.

\bibitem{Tsallis06c}
S. Umarov, C. Tsallis, M. Gell-Mann, and S. Steinberg, cond-mat/0606040.

\bibitem{Tsallis07}
A. Pluchino, A. Rapisarda, and C. Tsallis, \textit{E.P.L.} \textbf{80}, 26002
(2007).

\bibitem{Tsallis01}
V. Latora, A. Rapisarda, and C. Tsallis, \textit{Phys. Rev.} \textbf{E64},
056134 (2001).

\bibitem{Pluchino06}
A. Rapisarda and A. Pluchino, \textit{Europhys. News} \textbf{36}, 202 (2005). 

\bibitem{Bouchet06}
F. Bouchet, T. Dauxois, and S. Ruffo, \textit{Europhys. News} \textbf{37},
9 (2006).

\bibitem{Schehr07}
H.~J. Hilhorst and G. Schehr, \textit{J. Stat. Mech.} P06003 (2007).

\bibitem{Ellis85}
R.~S. Ellis, \textit{Entropy, Large Deviations and Statistical Mechanics},
(Springer, New York, 1985).

\bibitem{Botet02}
R. Botet and M. P{\l}oszajczak, \textit{Universal Fluctuations
--The Phenomenology of Hadronic Matter} (World Scientific, Singapore, 2002).

\bibitem{Taqqu75}
M.~S. Taqqu, \textit{Z. Wahrsch. Verw. Gebiete} \textbf{31}, 287 (1975).

\bibitem{Rosenblatt61}
M. Rosenblatt, \textit{Proceedings of the 4th Berkeley Symposium on
Math. Stat. and Prob.} (University of California, 1961).

\bibitem{Binder87}
K. Binder, in \textit{Phase Transitions and Critical Phenomena}, edited by
C. Domb and J.~L. Lebowitz (Academic, London, 1987), Vol.~8, p.~1,
and references therein.

\bibitem{Diehl87}
H.~W. Diehl, in \textit{Phase Transitions and Critical Phenomena}, edited by
C. Domb and J.~L. Lebowitz (Academic, London, 1987), Vol.~10, p.~76,
and references therein.

\bibitem{Wiese03}
K.J. Wiese, \textit{Ann. Henri Poincar\'e} \textbf{4}, 473 (2003).

\bibitem{Fisher75}
D.~R. Nelson and M.~E. Fisher, \textit{Ann. Phys} \textbf{91}, 226 (1975).

\bibitem{Droz78}
M. Droz and A. Malaspinas, \textit{J. Phys.} \textbf{C11}, 2729 (1978).

\bibitem{Lieb66}
E.~H. Lieb and D.~C. Mattis, \textit{Mathematical Physics in One Dimension:
Exactly Soluble Models of Interacting Particles} (Academic Press, London,
1966).

\bibitem{Naveau05}
P. Naveau, M. Nogaj, C. Ammann, P. Yiou, D. Cooley, and V. Jomelli,
\textit{C. R. Acad. Sci.} \textbf{337}, 1013 (2005).

\bibitem{Naveau06}
D. Cooley, P. Naveau, and V. Jomelli, \textit{Environmetrics} \textbf{17},
555 (2006)

\bibitem{Derrida81}
B. Derrida, \textit{Phys. Rev. Lett.} \textbf{45}, 79 (1980).

\bibitem{Derrida81b}
B. Derrida, \textit{Phys. Rev.} \textbf{B24}, 2613 (1981).

\bibitem{Carpentier}
D. Carpentier et P. Le Doussal, \textit{Phys. Rev.} \textbf{E63}, 026110
(2001).

\bibitem{Gyorgyi03}
G. Gy\"orgyi, P.~C.~W. Holdsworth, B. Portelli, and Z. R\'acz,
\textit{Phys. Rev.} \textbf{E68}, 056116 (2003).

\bibitem{Comtet04}
S.~N. Majumdar and A. Comtet, \textit{Phys. Rev. Lett.} \textbf{92}, 225501
(2004).

\bibitem{Comtet05}
S.~N. Majumdar and A. Comtet, \textit{J. Stat. Phys.} {bf 119}, 777 (2005).

\bibitem{Lee05}
D.~S. Lee, \textit{Phys. Rev. Lett.} \textbf{95}, 150601 (2005).

\bibitem{Gyorgyi07}
G. Gy\"orgyi, N.~R. Moloney, K. Ozog\'any, and Z. R\'acz,
\textit{Phys. Rev.} \textbf{E75}, 021123 (2007).

\bibitem{Johansson00}
K. Johansson, \textit{Commun. Math. Phys.} \textbf{209}, 437 (2000).

\bibitem{Krapivsky00}
S.~N. Majumdar and P.~L. Krapivsky, \textit{Phys. Rev.} \textbf{E62},
7735 (2000).

\bibitem{Dean01}
D.~S. Dean and S.~N. Majumdar, \textit{Phys. Rev.} \textbf{E64}, 046121 (2001).

\bibitem{Bena07}
I. Bena and S.~N. Majumdar, \textit{Phys. Rev.} \textbf{E75}, 051103 (2007).

\bibitem{Krug03}
J. Krug and C. Carl, \textit{Physica} \textbf{A318}, 137 (2003).

\bibitem{Schlather03}
M. Schlather and J. Tawn, \textit{Biometrika.} \textbf{90}, 139 (2003).

\bibitem{Cooley06}
D. Cooley, P. Naveau, and P. Poncet, ``Variograms for spatial max-stable
random fields'', in \textit{Dependence in Probability and Statistics}
(Springer, Berlin, 2006).

\bibitem{Burkhardt07}
T.~W. Burkhardt, G. Gy\"orgyi, N.~R. Moloney, and Z. R\'acz,
\textit{Phys. Rev.} \textbf{E76}, 041119 (2007).

\bibitem{BaldassarriTh}
A. Baldassarri, \textit{Statistics of Persistent Extreme Events},
Ph.D. Thesis, Univ. Paris XI, Paris, 1999.

\bibitem{Portelli02}
B. Portelli and P.~C.~W. Holdsworth, \textit{J. Phys.} \textbf{A35}, 1231
(2002).

\bibitem{PortelliTh}
B. Portelli, \textit{Fluctuations of Global Quantities in Correlated Systems},
Ph.D. Thesis, \'Ecole normale sup\'erieure de Lyon, Lyon, 2001.

\bibitem{Clusel04}
M. Clusel, J.-Y. Fortin and P.~C.~W. Holdsworth, \textit{Phys. Rev.}
\textbf{E70}, 046112 (2004).

\bibitem{Bertin05}
E. Bertin, \textit{Phys. Rev. Lett.} \textbf{95}, 170601 (2005).

\bibitem{BenArous06}
G. Ben Arous and D.~V. Voiculescu, \textit{Ann. Probab.} \textbf{34}, 1643
(2006).

\bibitem{Bertin06}
E. Bertin, \textit{J. Phys.} \textbf{A39}, 1539 (2006).

\bibitem{Panja04}
D. Panja, \textit{Phys. Rev.} \textbf{E70}, 036101 (2004).

\bibitem{Racz02}
T. Antal, M. Droz, G. Gy\"orgyi and Z. R\'acz, \textit{Phys. Rev.}
\textbf{E65}, 046140 (2002).

\bibitem{Racz03}
Z. R\'acz, in \textit{Proceedings of SPIE \textbf{5112}}, ed. M.~B. Weissman,
N.~E. Israeloff, A. Shulim Kogan, (2003), p. 248.

\bibitem{Frisch56}
H.~L. Frisch, Phys. Rev. \textbf{104}, 1 (1956).

\bibitem{Jepsen65}
D.~W. Jepsen, J. Math. Phys. \textbf{6}, 405 (1965).

\bibitem{Lebowitz67}
J.~L. Lebowitz and J.~K. Percus, Phys. Rev. \textbf{155}, 122 (1967).

\bibitem{Piasecki01}
J. Piasecki, J. Stat. Phys. \textbf{104}, 1145 (2001).

\bibitem{VandenBroeck02}
V. Balakrishnan, I. Bena, and C. Van den Broeck, Phys. Rev. E \textbf{65},
031102 (2002).

\bibitem{BCH08}
E. Bertin, M. Clusel, and P.~C.~W. Holdsworth, preprint arXiv:0803.4149,
to appear in \textit{J. Stat. Mech.} (2008).

\bibitem{Archambault97}
P. Archambault, S.~T. Bramwell and P.~C.~W. Holdsworth, \textit{J. Phys.}
\textbf{A30}, 8363 (1997).

\bibitem{EW82}
S.~F. Edwards and D.~R. Wilkinson, \textit{Proc. Roc. Soc. London}
\textbf{A381}, 17 (1982).

\bibitem{Jensen01}
K. Dahlstedt and H.~J. Jensen, \textit{J. Phys.} \textbf{A34}, 11193 (2001).

\bibitem{Watkins02}
N.~W. Watkins, S.~C. Chapman and G. Rowlands, \textit{Phys. Rev. Lett.}
\textbf{89}, 208901 (2002).

\bibitem{Bramwell02}
S.~T. Bramwell \textit{et al.}, \textit{Phys. Rev. Lett.} \textbf{89}, 208902
(2002).

\bibitem{Banks05}
S.~T. Banks and S.~T. Bramwell, \textit{J. Phys.} \textbf{A38}, 5603 (2005).

\bibitem{Mack05}
G. Mack, G. Palma and L. Vergara, \textit{Phys. Rev.} \textbf{E72}, 026119
(2005).

\bibitem{Clusel06}
M. Clusel, J.-Y. Fortin and P.~C.~W. Holdsworth, \textit{Europhys. Lett.}
\textbf{76}, 1008 (2006).

\bibitem{Goldenfeld92}
N. Goldenfeld, \textit{Lectures on Phase Transitions and the Renormalization
Group Approach} (Perseus Books, Reading, 1992).

\bibitem{CluselTh}
M. Clusel, \textit{Some Aspects of Statistical Physics of Correlated Systems},
Ph.D. Thesis, \'Ecole normale sup\'erieure de Lyon, Lyon (2005).

\bibitem{Portelli01}
B. Portelli, P.~C.~W. Holdsworth, M. Sellitto, and S.~T. Bramwell,
\textit{Phys. Rev.} \textbf{E64}, 036111 (2001).

\bibitem{Leung82}
K.~M. Leung, \textit{Phys. Rev.} \textbf{B35}, 226 (1982).

\bibitem{BenArous05}
G. Ben Arous, L.~B. Bogachev, S.~A. Molchanov, \textit{Probab. Theory Relat.
Fields} \textbf{132}, 579 (2005).



\end{thebibliography}
\end{document}